\documentclass[aps,showpacs,showkeys,preprint,superscriptaddress]{revtex4}
\usepackage[caption = false]{subfig}
\usepackage[final]{graphicx}
\usepackage[usenames, dvipsnames]{color}
\usepackage{float}
\usepackage{hyperref}
\usepackage{bm}
\usepackage{amsmath}
\usepackage{txfonts}
\usepackage{epstopdf}
\begin{document}
\title{Transient and pre-transient stages in field induced phase transition of vacuum state}
\author{Chitradip Banerjee\footnote{Corresponding author.\\E-mail address: cbanerjee@rrcat.gov.in (C. Banerjee).}}
\author{Manoranjan P. Singh}
\affiliation{Theory and Simulations Lab, HRDS, Raja Ramanna Centre for Advanced
Technology,  Indore-452013, India}
\affiliation{Homi Bhabha National Institute, Training School Complex, Anushakti Nagar, Mumbai 400094, India}
\begin{abstract}
Evolution of modulus and phase of complex order parameter associated with field induced phase transition (FIPT) of the vacuum state interacting with time dependent Sauter pulse is studied to analyse different evolution stages of the order parameter e.g., quasi electron positron plasma (QEPP), transient, and residual electron-positron plasma (REPP) stages. By revisiting FIPT in presence of single-sheeted and multi-sheeted pulses, we attribute the transient stage  to the nonlinear coupling in the differential equations governing the dynamics of the phase and the modulus of the order parameter. Appearance of the rapid oscillations in the modulus is shown to be associated with the abrupt change in the phase of the order parameter in the transient stage. FIPT is also studied for multi-sheeted Sauter pulse with linear and quadratic frequency chirp.  QEEP stage is found to show complex dynamical behaviour with fast and irregular oscillations due to the frequency chirp.  The formation of pre-transient region due to the quadratic frequency chirping is observed in the accelerating part of the QEPP stage before the electric field attains the maximum value.  As the quadratic chirp is increased the pre-transient and transient stages move closer to the electric field maximum which leads to decrease in temporal separation between the two stages. Early appearance of the transient stage and hence of the following REPP stage results in the enhancement of pair production rate.
\end{abstract}
\maketitle
\section{introduction}
The electron-positron pair production from an unstable vacuum in the presence of a uniform electric field, widely known as the Schwinger mechanism, \cite{Heisenberg,Sauter,PhysRev.82.664,nikishov1970pair,NarozhnyiNikishov1970} is one of the intriguing phenomenon in the nonperturbative regime of quantum electrodynamics (QED). This process is suppressed exponentially for the electric field strength $E \ll E_S$, the critical electric field of QED ($E_S = m^2c^3/{e\hbar} = 1.38\times 10^{18}$ V/m) \cite{Sauter} and the probability of pair production in the presence of a constant electric field $E$ is given by $P_{e^+e^-} \approx \exp(-\pi E_S/E)$. As $E\ll E_S$ for state of the art present day ultraintense lasers and other light sources, this fundamental prediction of QED could not be tested experimentally. However, recent advances in laser technology, specially the use of chirp pulse amplification method, promise to fast narrow the gap \cite{PhysRevSTAB.5.031301}. The European Extreme Light Infrastructure for Nuclear Physics (ELI-NP) is planing to build a $10$~PW pulsed laser to achieve intensities $I \sim 10^{23}~\textnormal{W/cm}^2$ for the first time for investigating new physical phenomena at the interfaces of plasma, nuclear and particle physics \cite{Heinzl2009EPJD,Dunne2009EPJD}. The electric field at the laser focus will have a maximum value of $10^{15}$ V/m at such intensities. In the ELI-NP experimental area E6, there is proposal to study radiation reactions, strong field QED effects and the resulting production of ultrabright gamma rays which could be used for nuclear activation. The construction of X-ray free electron (XFEL) based on the principle of self amplified spontaneous emission (SASE)  is under way at DESY, Hamburg  \cite{PhysRevLettDESYXFEL}. In a landmark experiment E$144$ at Stanford Linac Acceleration centre (SLAC) in 1997 it was possible to observe non-linear QED processes like non-linear Compton scattering and stimulated pair production in the collision of a $46.6$ GeV electron beam with terawatt photon pulses of $1053$ nm and $527$ nm in multi-photon regime at laser intensity $I = 10^{22}~\textnormal{W/cm}^2$ \cite{CbulaPhysRevLett.76.3116,DbrukePhysRevLett.79.1626}. Although these processes pertain to the perturbative regime of QED, the successful experimental realizations thereof raise the hope for the experimental verification of the Schwinger mechanism in coming decades.  

The original formalism of the Schwinger mechanism was developed for the electric field constant in space and time. It was generalized subsequently for time varying field \cite{PhysRevD.2.1191Itzykson} and is applied to study pair production by ultraintense and ultrashort laser pulses on the consideration that the length and time scales of the variation in the field is much less than the Compton length and time. As the pulse duration of some of the proposed light sources e.g., lasers based on higher order harmonics, is expected to be of the order of attoseconds, one needs to look beyond the Schwinger formalism. Furthermore, for the dynamical studies  like longitudinal momentum spectrum of created particles one has to solve the underlying Dirac equation for the fermionic system (Klein-Gordon equation for scalar field) interacting with the external electromagnetic field  \cite{PhysRevD.73SPKim2006,PhysRevD.82Dumlu,PhysRevDDumlu2011,PhysRevD2011Fedotov,PhysRevDSPKim2002,PhysRevD2011KimQVE,PhysRevDSPKim2008,PhysRevDSPKim2009}. In order to have analytically tractable dynamical description for the system of unstable vacuum in the presence of the external field two broad classes of theoretical formulations based on the spatiotemporal inhomogeneity of the external field were developed \cite{PRDGitmanVacuumInstability}. These are termed as the kinetic equation in Wigner \cite{QGPWigner,PRDHebenstreitWignerKinetic,PhysRevDBialynickiBirulaWigner} and in the quasi-particle \cite{PhysRevD1999Schmidt,Schmidth1998IJMPE,DynamiSchwingerBlaschke} representations. For the one dimensional spatially uniform time dependent electric, the equivalence between these representations of the kinetic equation was shown in \cite{2011PhDTHebenstreit}.
We use here the quasi-particle representation of the quantum kinetic equation (QKE) formalism  \cite{PhysRevDKluger1998,PhysRevD1999Schmidt,PhysRevLett.67.2427Kluger,PhysRevD1994JRau,Schmidth1998IJMPE,Filatov2008} for the evolution of the quasi-particle vacuum in the presence of a time dependent electric field in the mean field approximation wherein the collisional effects of the created particles and back reaction force on the external electric field are neglected.

The production of particle-antiparticle pairs from the vacuum fluctuation in a time-dependent electric field $E(t)$ was seen as a field induced phase transition (FIPT) via the $t$-non invariant vacuum state because of the non-stationary Hamiltonian \cite{DynamiSchwingerBlaschke}. Here, the spontaneous symmetry breaking of the vacuum state takes place under time inversion and consequently electron-positron pairs are generated which are the massive analogue of Goldstone bosons \cite{SColeman}. In order to quantify this symmetry breaking one defines a complex order parameter $\Phi(\textbf{p},t) = 2\langle 0_{in}|a^{\dagger}_{\textbf{p}}(t)b^{\dagger}_{-\textbf{p}}(t)|0_{in}\rangle = |\Phi(\textbf{p},t)|\exp(i\psi(\textbf{p},t))$ \cite{DynamiSchwingerBlaschke,PhysRevD1999Schmidt,Smolyansky2012TimeReversalSymmetry} where $a^{\dagger}_{\textbf{p}}(t)$, $b^{\dagger}_{-\textbf{p}}(t)$ are creation operators of particle and antiparticle with momentum $\pm \textbf{p}$, respectively, in the quasi-particle representation in the time dependent basis. FIPT was studied for the single and multi-sheeted electric pulses \cite{DynamiSchwingerBlaschke,PhysRevD1999Schmidt,Smolyansky2012TimeReversalSymmetry}. It was shown that the evolution of the modulus of order parameter $|\Phi(\textbf{p},t)|$ brings out three distinct stages/ phases namely the quasi-electron positron plasma (QEPP) stage, the transient stage and the final residual electron positron plasma stage (REPP). The effect of subcycle field oscillations on these stages was also  studied for different longitudinal momentum values \cite{Smolyansky2012TimeReversalSymmetry,Smolyansky2017FieldPhaseTrans}. However the evolution of the phase of the complex order parameter $\psi(\textbf{p},t)$, to the best of our knowledge, has not been studied so far.

In this Letter, we study the evolution of the modulus and the phase of $\Phi(\textbf{p},t)$ and analyse their interrelation. Frequency chirp is an essential and integral part of ultrashort laser pulses. The frequency chirping changes the time period of subcycle oscillations of the external electric field which further induces complexity in the evolution of the order parameter. We analyse this complexity as a function of linear and quadratic frequency chirp parameters.

The rest of the Letter is organized as follows. In Sec.~\ref{theory} we describe the basic equations governing the evolution of the modulus and the phase of $\Phi(\textbf{p},t)$. Results are discussed in Sec.~\ref{results}. The Letter is concluded in Sec.~\ref{concl}.
\section{theory}\label{theory}
Using the quantum kinetic equations in the form of 3-coupled ordinary differential equations \cite{PhysRevD1999Schmidt,PhysRevDKluger1998,BlaschkeCPP,Filatov2008,DynamiSchwingerBlaschke} for the single particle distribution function $f(\textbf{p},t)=\langle 0_{in}|a^{\dagger}_{\textbf{p}}(t)a_{\textbf{p}}(t)|0_{in}\rangle = \langle 0_{in}|b^{\dagger}_{-\textbf{p}}(t)b_{-\textbf{p}}(t)|0_{in}\rangle$ and the real and imaginary parts of the order parameter $u(\textbf{p},t)=|\Phi(\textbf{p},t)|\cos{\psi(\textbf{p},t)}$ and $v(\textbf{p},t) = |\Phi(\textbf{p},t)|\sin{\psi(\textbf{p},t)}$ respectively, 
\begin{equation}\label{Set_ODE_KE}
 \begin{split}
 \frac{df(\textbf{p},t)}{dt} = \frac{eE(t)\epsilon_{\perp}}{2\omega^2(\textbf{p},t)}u(\textbf{p},t),\\
 \frac{du(\textbf{p},t)}{dt} =\frac{eE(t)\epsilon_{\perp}}{\omega^2(\textbf{p},t)}[1-2f(\textbf{p},t)]-2\omega(\textbf{p},t)v(\textbf{p},t),\\
 \frac{dv(\textbf{p},t)}{dt} = 2\omega(\textbf{p},t)u(\textbf{p},t),
 \end{split}
 \end{equation}
and using the first integral of motion $(1-2f(\textbf{p},t))^2+|\Phi(\textbf{p},t)|^2 = 1$, we get the following nonlinear coupled differential equations for the evolution of the modulus $|\Phi(\textbf{p},t)|$ and the phase $\psi(\textbf{p},t)$  of the order parameter
\begin{equation}\label{Ord_phase_KE}
\begin{split}
\frac{d|\Phi(\textbf{p},t)|}{dt} = \frac{eE(t)\epsilon_{\perp}}{\omega^2(\textbf{p},t)}\cos{\psi(\textbf{p},t)}\sqrt{1-|\Phi(\textbf{p},t)|^2},\\
\frac{d\psi(\textbf{p},t)}{dt} = 2\omega(\textbf{p},t)- \frac{eE(t)\epsilon_{\perp}}{\omega^2(\textbf{p},t)}\sin{\psi(\textbf{p},t)}\frac{\sqrt{1-|\Phi(\textbf{p},t)|^2}}{|\Phi(\textbf{p},t)|}.
\end{split}
\end{equation}
Here $u(\textbf{p},t)$ and $v(\textbf{p},t)$ govern the vacuum polarization and the counter process of pair production i.e., pair annihilation, respectively. The terms $\omega(\textbf{p},t) = \sqrt{m^2+p^2_{\perp}+P_3^2(t)}$ and $P_3(t) = p_3-eA(t)$ are the quasi-energy and the longitudinal quasi-momentum respectively of the quasi-particle. The particle acceleration is governed by $dP_3(t)/{dt} = eE(t)$ in the presence of the time dependent electric field $E(t)$; $e$ is the electronic charge; $\epsilon_{\perp} = \sqrt{m^2+\textbf{p}_{\perp}^2}$ is the transverse energy.

The electric field in this study is  taken as the multi-sheeted Sauter pulse which is considered to describe well the resultant field of counter propagating ultrashort laser pulses in the focal region. 
\begin{equation} 
E(t) = E_0\cosh^{-2}(t/\tau)\cos(\alpha t^3+\beta t^2+\omega_0 t),
\end{equation}
where $\beta$ and $\alpha$ are the linear and quadratic frequency chirp parameters respectively; $\omega_0$ is the central frequency of the laser electric field oscillation with $\tau$ being the total pulse length. single-sheeted Sauter pulse corresponds to $\alpha = \beta = \omega_0 = 0$. 

\section{Results}\label{results}
We solve Eq.~\ref{Ord_phase_KE} numerically for the evolution of $|\Phi(\textbf{p},t)|$ and $\psi(\textbf{p},t)$  with the initial condition $|\Phi(\textbf{p},t\rightarrow\infty)| = 0$. The phase $\psi(\textbf{p},t)$ is  defined only up to an arbitrary additive constant. For definiteness we begin at $t = -10\tau$ with $|\Phi(\textbf{p},t = -10\tau)| = 10^{-16}$ and $\psi(\textbf{p},t = -10\tau) = \pi/4$ so as to have $u = v = 10^{-16}$ initially. 

\begin{figure}[h]
\begin{center}
{\includegraphics[width = 2.5in]{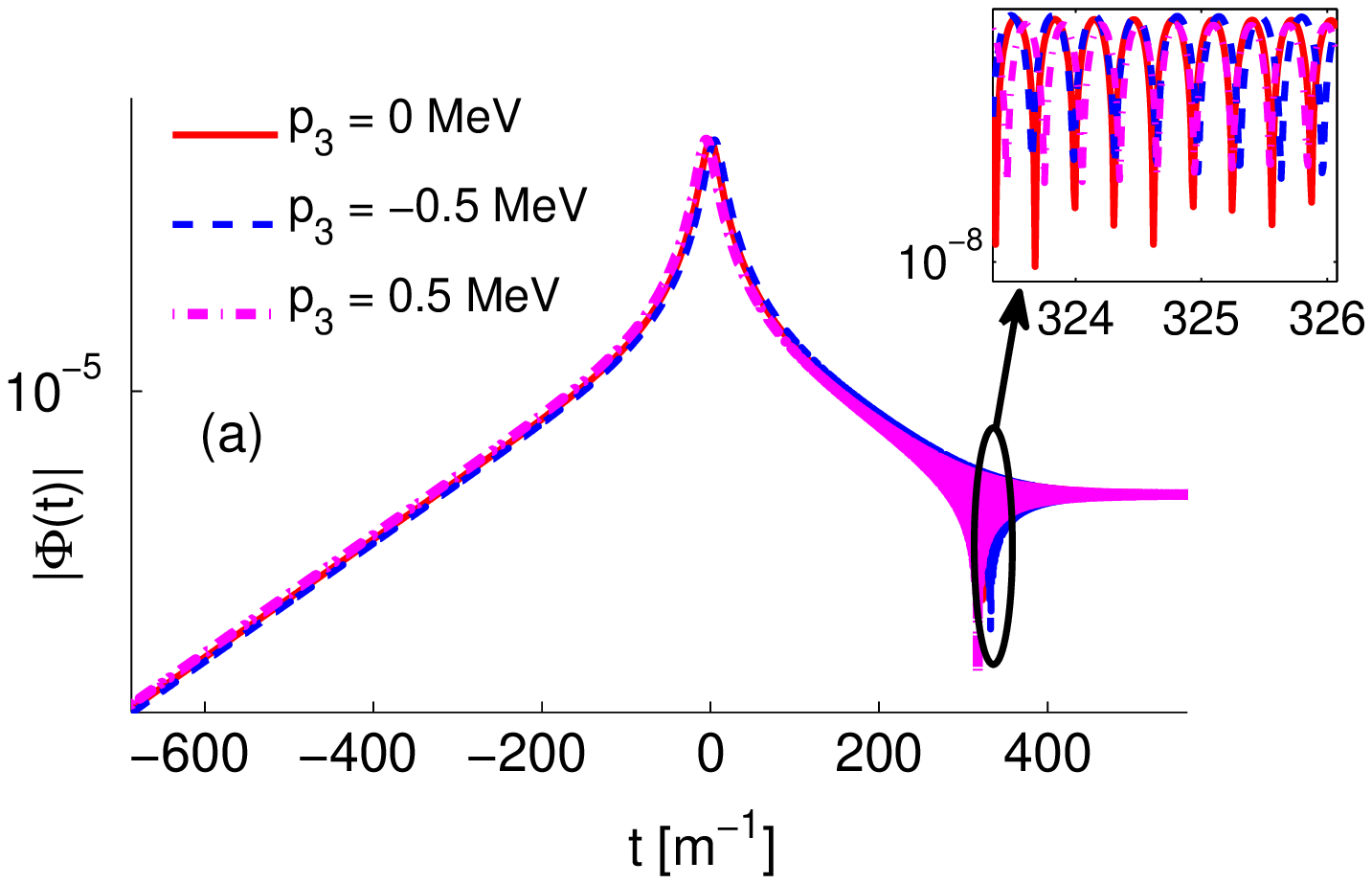}~~ 
\includegraphics[width = 2.5in]{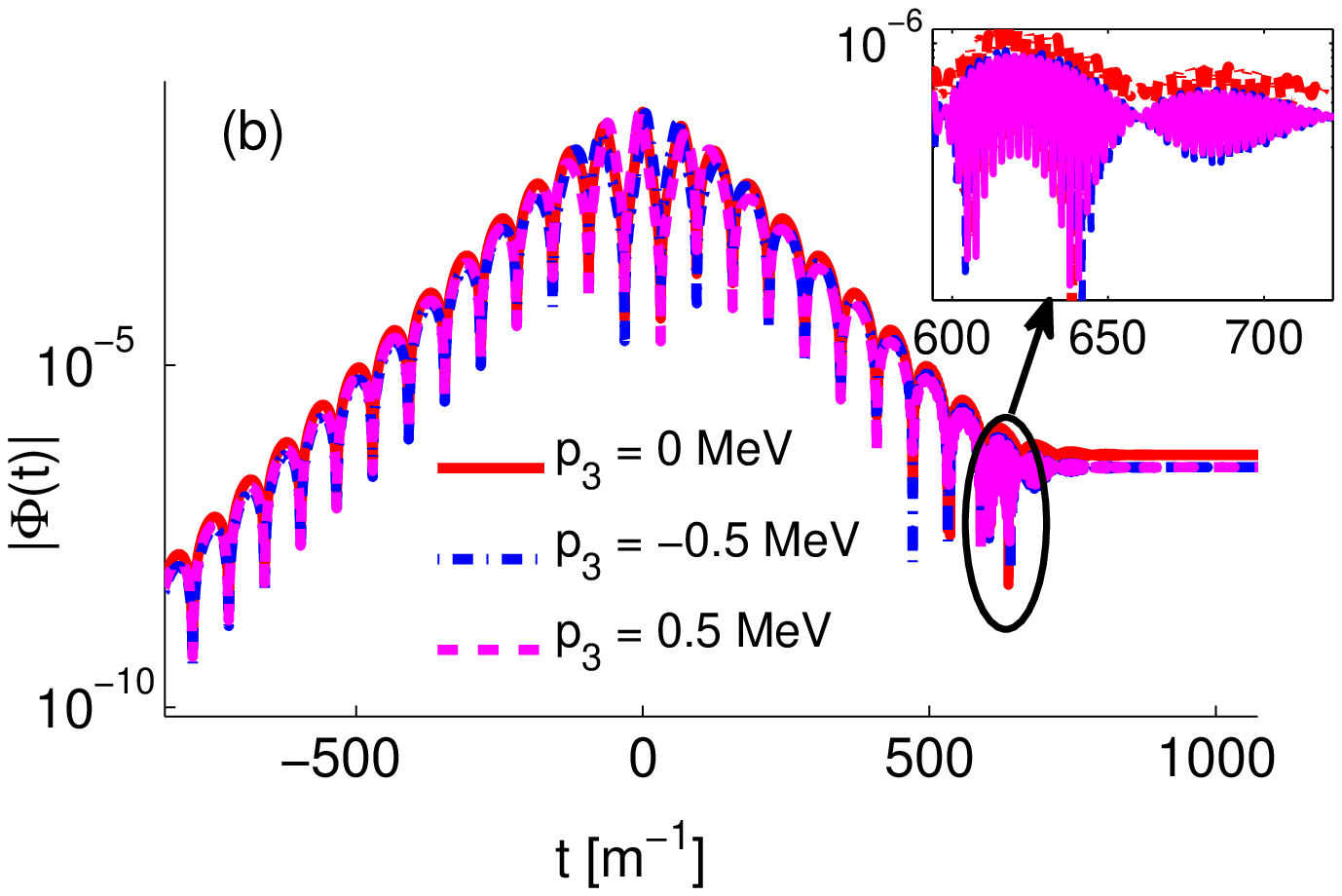}\\
\includegraphics[width = 2.5in]{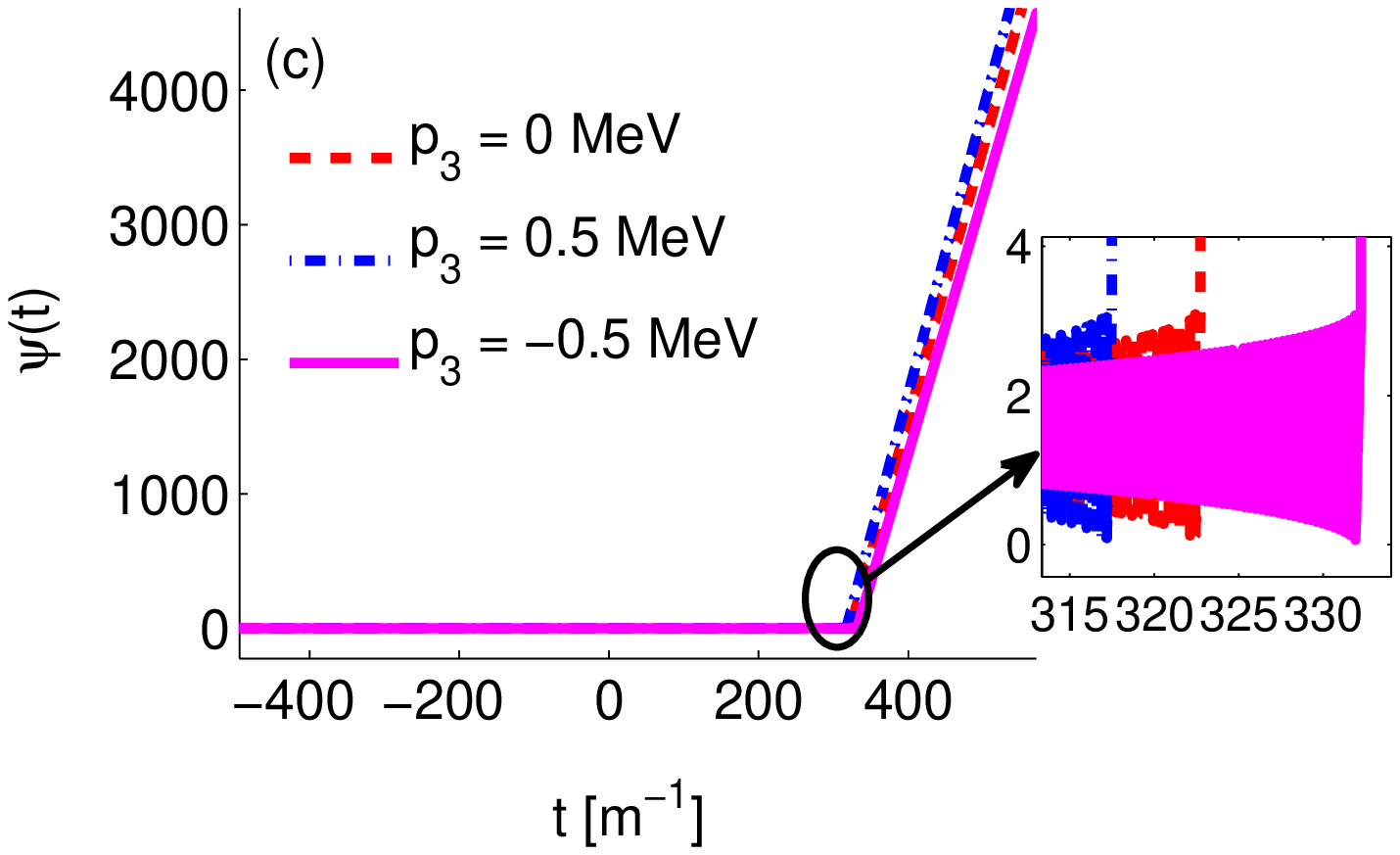}~~
\includegraphics[width = 2.5in]{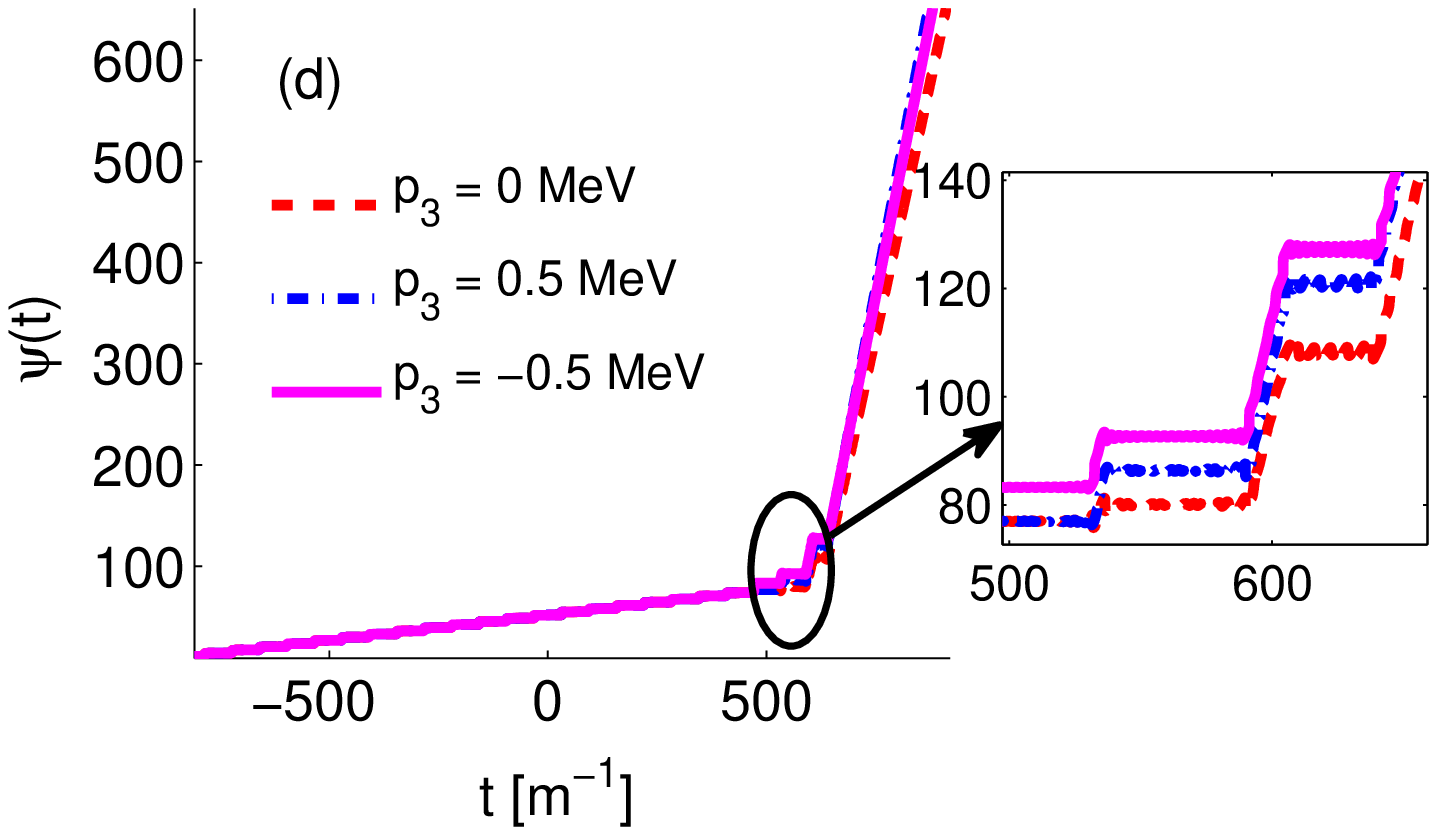}}
\caption{Evolution of the modulus $|\Phi(\textbf{p}, t)|$ and the phase $\psi(\textbf{p},t)$ of the order parameter -- (a) and (c) for single-sheeted Sauter pulse;(b) and (d) for multi-sheeted ($\omega_0\tau = 5$) Sauter pulse without frequency chirping ($\alpha = \beta = 0$) for the longitudinal momentum values $p_3 = 0 \textnormal{MeV}, \mp 0.5 \textnormal{MeV}$. All the units are taken in electron mass unit and the transverse momentum $\textbf{p}_{\perp} =0$. The field parameters are $E_0 = 0.1$, and $\tau = 100$. The insets show magnified view of the evolution in the transient region.}
\label{F_t_Sau_OmTau_05}
\end{center}
\end{figure}

As mentioned before, the complete evolution of the modulus of the order parameter $|\Phi(\textbf{p},t)|$ was shown to go through three distinct stages namely the initial QEPP stage, the transient stage, and the final REPP stage of the created electron-positron pairs by the external time dependent electric field  of single-sheeted Sauter pulse and multi-sheeted Gaussian pulse \cite{Smolyansky2012TimeReversalSymmetry,Smolyansky2017FieldPhaseTrans}. However, it is not clear what is the origin of the transient stage. Furthermore, the evolution of the phase $\psi(\textbf{p},t)$ has not been studied so far.  We, therefore revisit these cases. $|\Phi(\textbf{p},t)|$ and $\psi(\textbf{p},t)$ are plotted as function of time for single and multi-sheeted Sauter pulses without any frequency chirp ($\omega_0\tau = 5, \alpha = \beta = 0$) in Fig.~\ref{F_t_Sau_OmTau_05}. It is seen in Fig.~\ref{F_t_Sau_OmTau_05}(a) that $|\Phi(\textbf{p},t)|$ increases monotonically with time in QEPP region for the single-sheeted  Sauter pulse up to $t = 0$ at which electric field reaches its maximum value and thereafter $|\Phi(\textbf{p},t)|$ decreases. $|\Phi(\textbf{p},t)|$, after decreasing to a certain value, shows rapid oscillations and thus undergoes transition from QEPP to a transient region before settling down to REPP state as a consequence of FIPT wherein $|\Phi(\textbf{p},t)|$ reaches a constant value different from zero. The phase $\psi(\textbf{p},t)$ remains almost constant before increasing rapidly about the time the transient stage in the evolution of $|\Phi(\textbf{p},t)|$ appears, see Fig.~\ref{F_t_Sau_OmTau_05}(c). For the multi-sheeted Sauter pulse the evolution of $|\Phi(\textbf{p},t)|$, in Fig.~\ref{F_t_Sau_OmTau_05}(b)  shows periodic oscillations corresponding to the subcycle structure of the electric field in the QEPP region. The transient region  appears later and becomes elongated before it reaches final REPP state. $\psi(\textbf{p}, t)$ is found to increase slowly in the QEPP region and then rapidly at the onset of the transient stage, note the abrupt change of slope in the linear growth of $\psi(\textbf{p}, t)$ in Fig.~\ref{F_t_Sau_OmTau_05}(d). The blow up of the evolution of $\psi(\textbf{p}, t)$ in the transient stage shows a stair-case like structure. 

\begin{figure}[h]
\begin{center}
{\includegraphics[width = 2.0in]{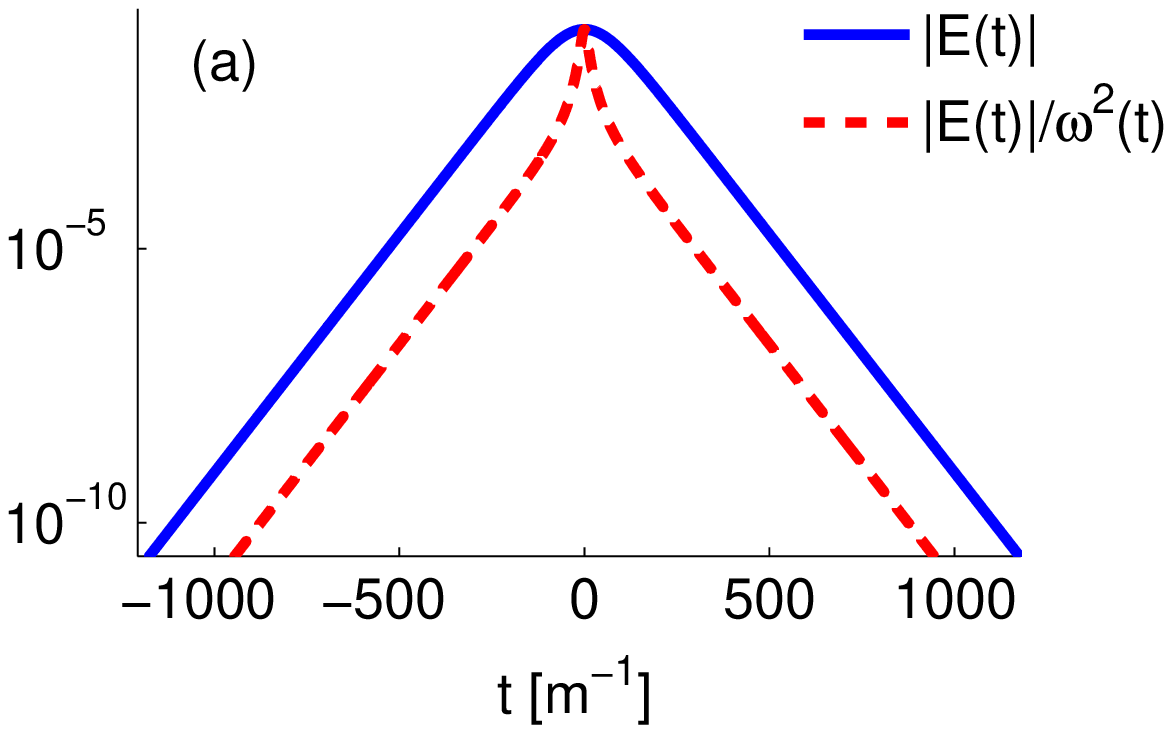}~
\includegraphics[width = 2.0in]{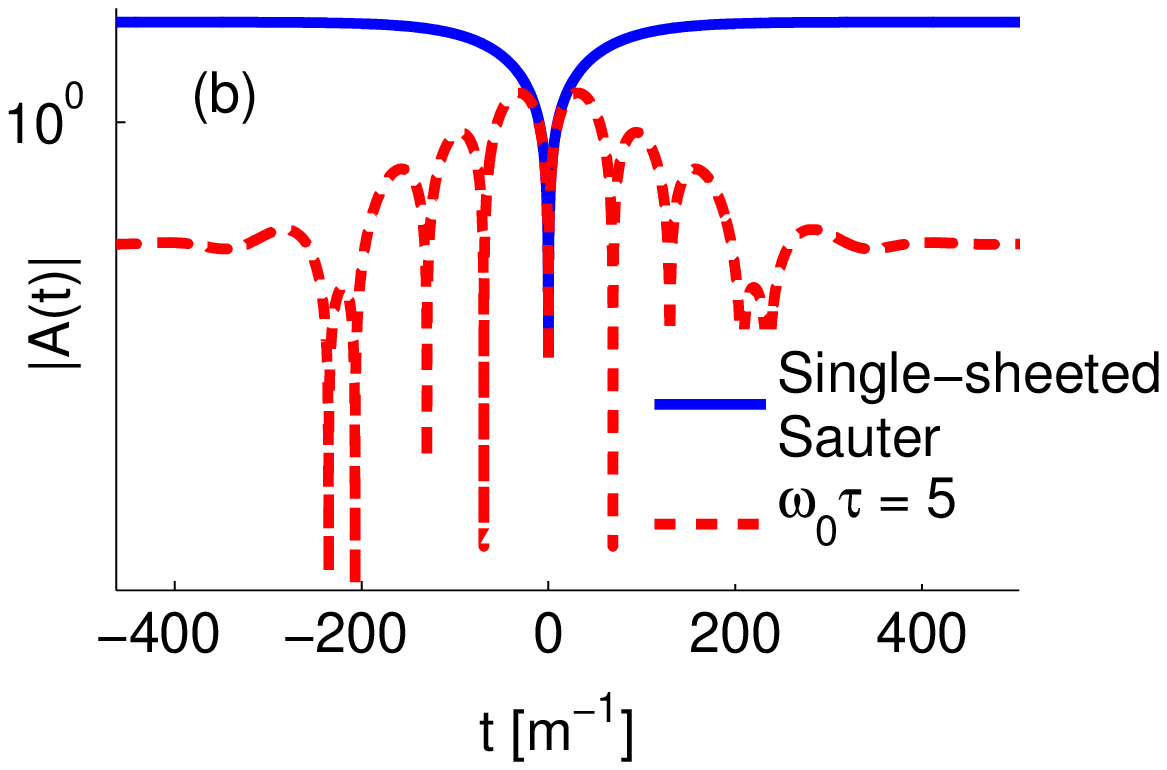}~
\includegraphics[width = 2.0in]{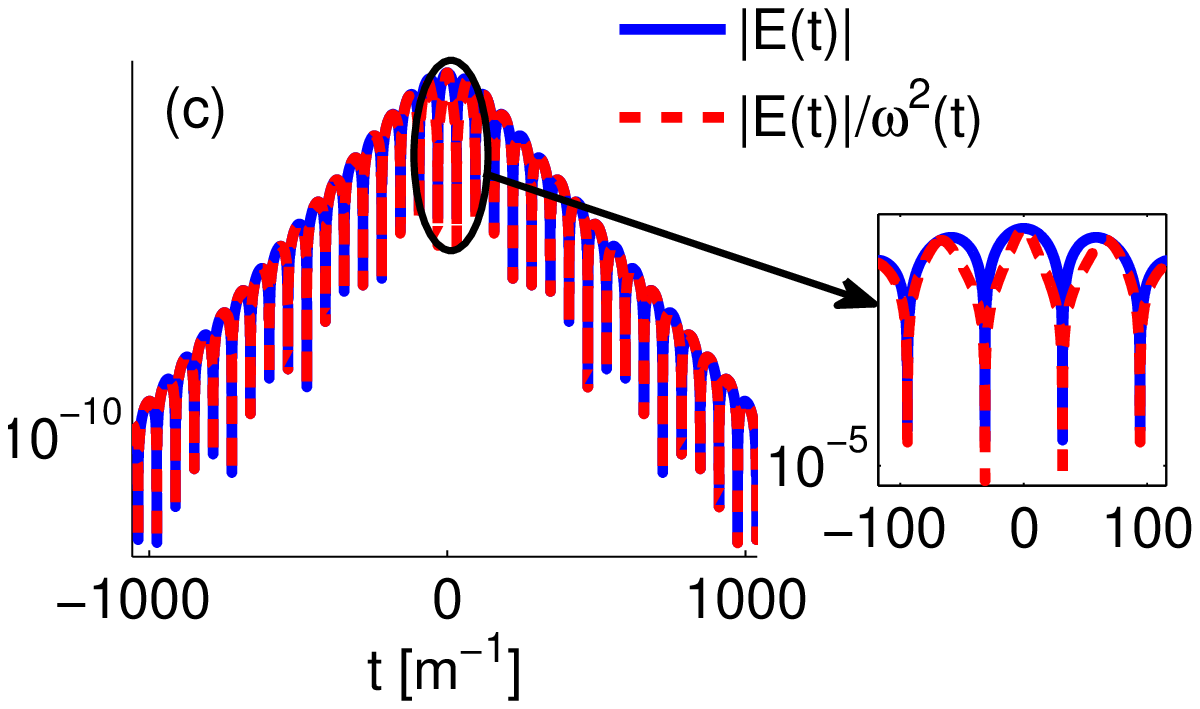}\\
\includegraphics[width = 3.0in]{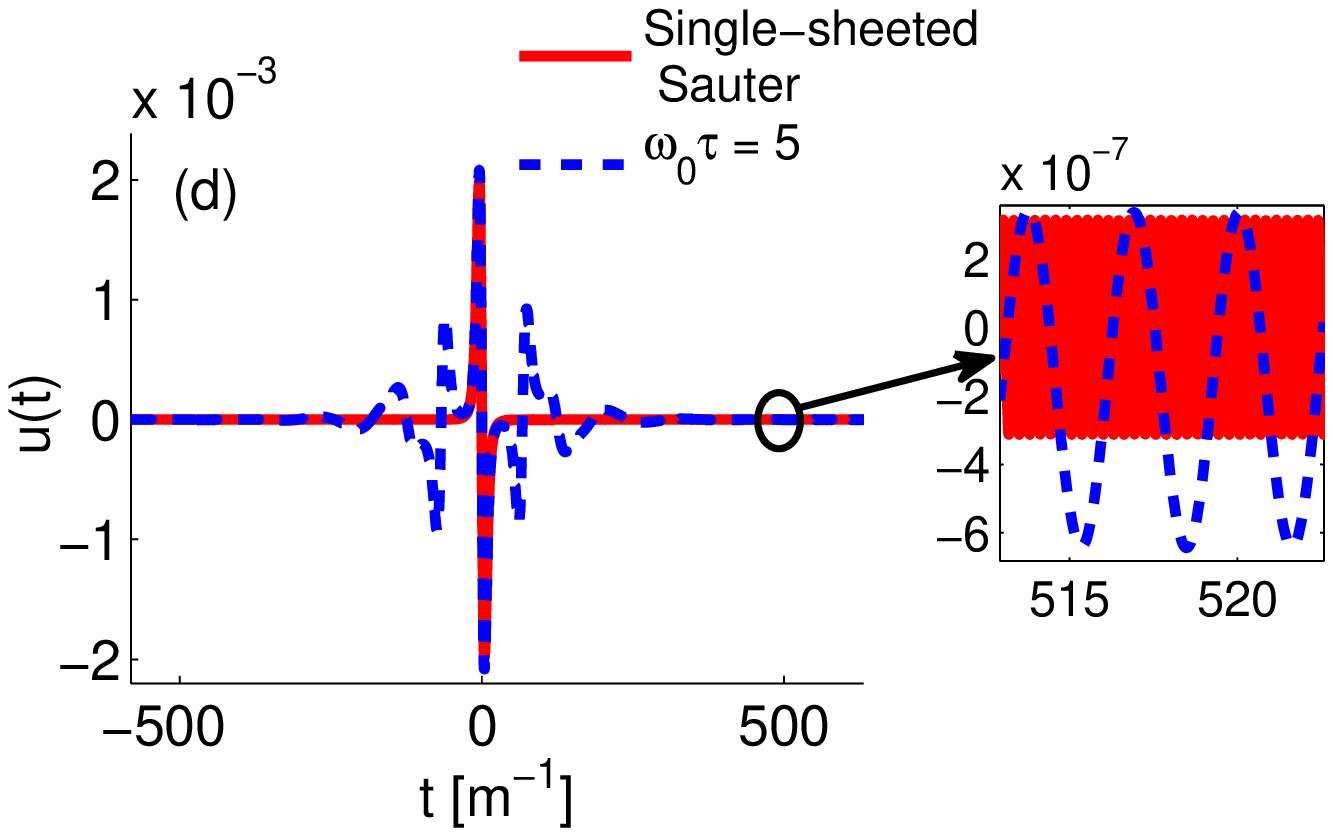}~~
\includegraphics[width = 3.0in]{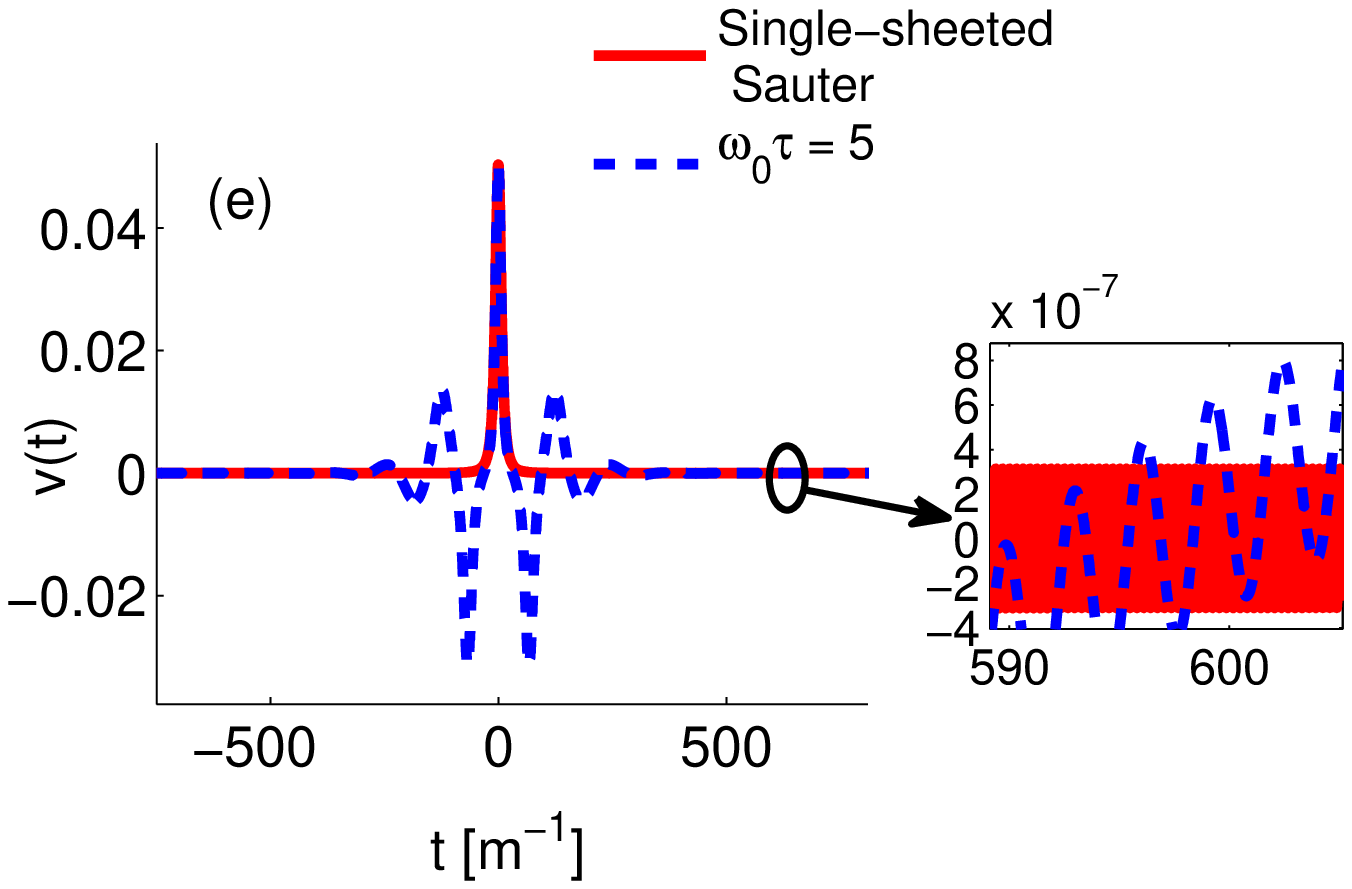}}
\caption{(a) and (c) $E(t)$, $|E(t)|/\omega^2(\textbf{p},t)$  for  single-sheeted and multi-sheeted ($\omega_0\tau = 5)$ Sauter pulses respectively. (b) $A(t)$ for single and multi-sheeted pulses. (d) and (e) Evolutions of  $u(\textbf{p}, t)$  and  $v(\textbf{p}, t)$, respectively  for single and multi-sheeted pulses, with insets showing the magnified view of evolution in the transient stage. All the units are taken in electron mass unit. The field parameters are $E_0 = 0.1$, $\tau = 100$, the central frequency of the pulse $\omega_0 = 0.05$, and linear and quadratic frequency chirp parameters $\beta = \alpha = 0$. Transverse and longitudinal momenta are taken to be zero.}
\label{E_A_t_Sau_OmTau5_b0c0}
\end{center}
\end{figure}

\begin{figure}[h]
\begin{center}
{\includegraphics[width = 2.0in]{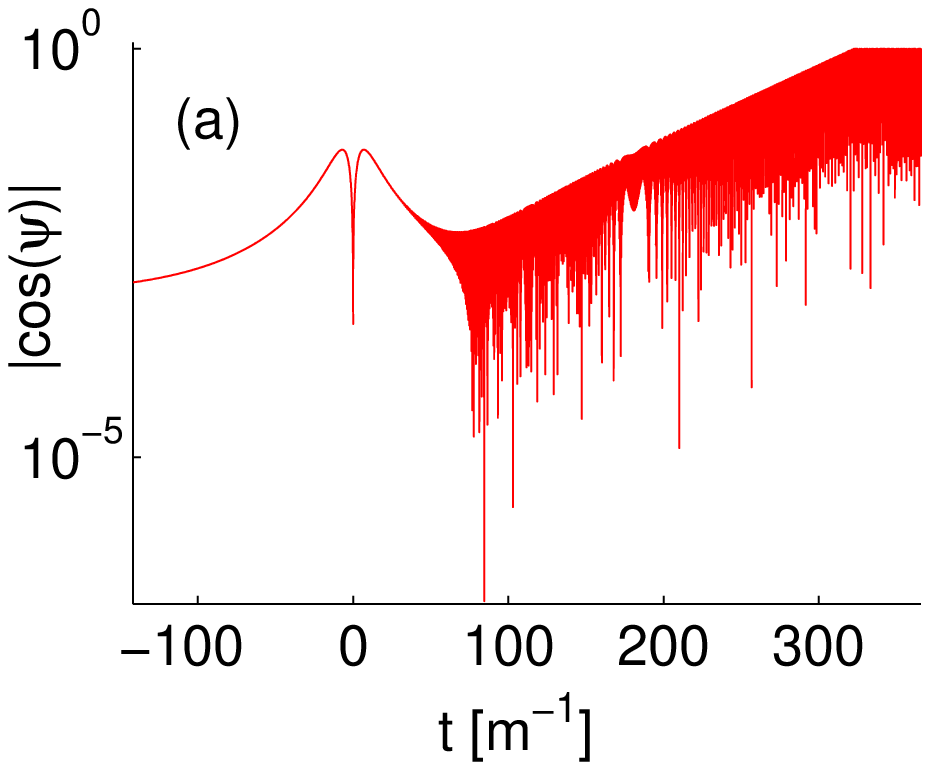}~
\includegraphics[width = 2.0in]{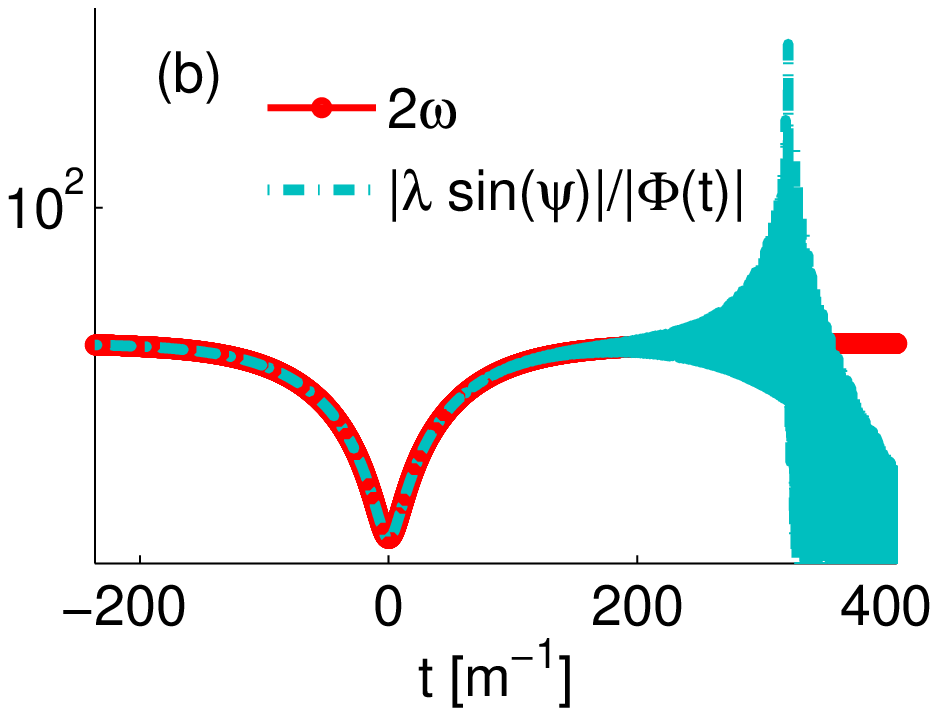}\\
\includegraphics[width = 2.0in]{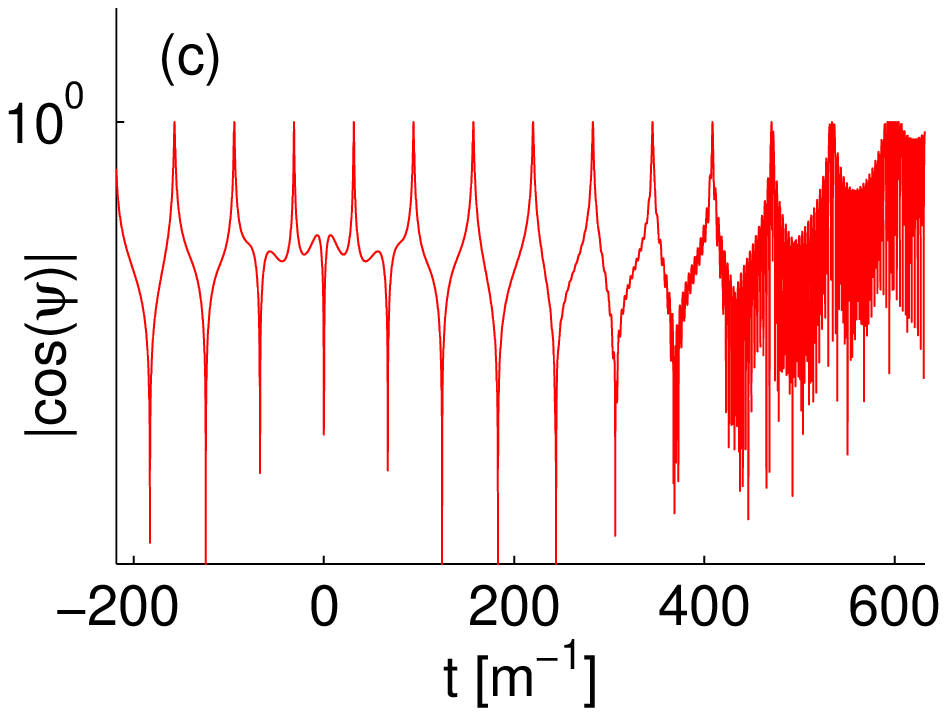}
\includegraphics[width = 2.0in]{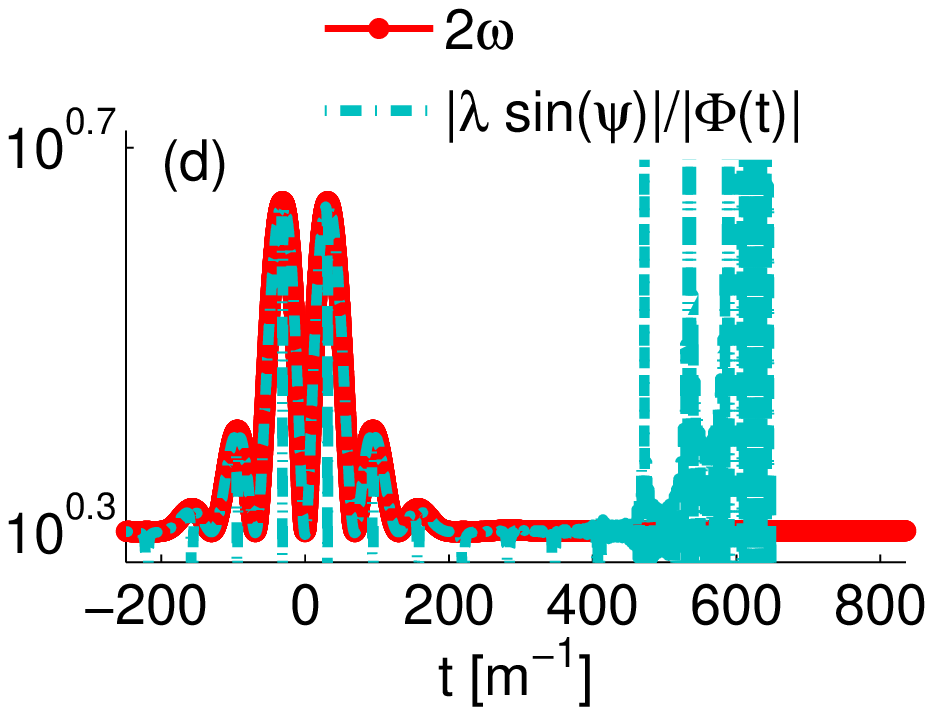}}
\caption{(a) and (c) $|\cos{\psi(\textbf{p}, t)}|$ for single-sheeted and multi-sheeted Sauter pulses respectively. (b) and (d) $2\omega(\textbf{p}, t)$ and $|\lambda\sin{\psi(\textbf{p}, t)}|/|\Phi(\textbf{p}, t)|$ for single-sheeted and multi-sheeted Sauter pulses respectively. Here, $\lambda = E(t)/\omega^2(\textbf{p}, t)$. The longitudinal and transverse momenta are $p_3 = p_{\perp} = 0$. The field parameters are $E_0 = 0.1$, $\tau = 100$.}
\label{coslambdaOmT5}
\end{center}
\end{figure}

As seen in Eq.~\ref{Ord_phase_KE} the evolution of $|\Phi(\textbf{p},t)|$ is governed by the temporal profiles of the electric field $E(t)$ and the corresponding vector potential $A(t)$ through the ratio $ E(t)/\omega^2(\textbf{p},t)$ (note that $\epsilon_{\perp} = e = 1$) and also by the phase term $\cos{\psi(\textbf{p},t)}$. As $\psi(\textbf{p},t)$ remains nearly constant, the QEPP stage is largely controlled by $E(t)/\omega^2(\textbf{p},t)$ as seen the Fig.~\ref{F_t_Sau_OmTau_05}(a). For the single-sheeted pulse as shown in Fig.~\ref{E_A_t_Sau_OmTau5_b0c0}(a) the electric field profile is smooth having its maximum at $t = 0$ while the vector potential is large in magnitude on the either side of the electric field maximum resulting in a sharper temporal profile of $E(t)/\omega^2(\textbf{p},t)$. In Refs. \cite{BlaschkeCPP,Smolyansky2012TimeReversalSymmetry,Smolyansky2017FieldPhaseTrans} the temporal profile of $|\Phi(\textbf{p},t)|$ in the QEPP stage was compared to that of $|E(t)|$. However, the much sharper profile of  $|\Phi(\textbf{p},t)|$, particularly near the centre of the pulse, and the faster decay thereof in the tail regions is better explained by $|E(t)|/\omega^2(\textbf{p},t)$ than  $|E(t)|$, see  Fig.~\ref{E_A_t_Sau_OmTau5_b0c0}(a). The formation of the transient region takes place  because of the sudden rapid increase in the value of the phase $\psi(\textbf{p},t)$ which makes $\cos{\psi(\textbf{p},t)}$ (on the right hand side of Eq.~\ref{Ord_phase_KE}) and hence $|\Phi(\textbf{p},t)|$ oscillate rapidly, see Fig.~\ref{coslambdaOmT5}(a).  Once the electric field gets vanishingly small, ${d|\Phi(\textbf{p},t)|}/{dt} = 0$ and we have a constant value of $|\Phi(\textbf{p},t)|$ in the REPP region. In this region $d\psi(\textbf{p},t)/dt = 2 \omega(\textbf{p}, t)$ and hence $\Phi(\textbf{p},t) \sim \Phi_R e^{2 i \omega t}$ where $\Phi_R$ is the constant value of $\Phi(\textbf{p},t)$. $\Phi_R$  being largest for $p_3=0$ and the same for $p_3=\pm 0.5$ MeV is consistent with the well known fact that the asymptotic momentum spectrum of pairs created by the single-sheeted pulse is centred at $p_3=0$ and is symmetric about it \cite{PhysRevD.82.105026Hebenstreit}.  At this point it is worthwhile noting that in the QEPP stage, for $t \ge 0$ and in the transient stage $|\Phi(\textbf{p},t)|$ is largest for $p_3=-0.5$ MeV  which suggests that the (quasi) particle momentum spectrum should be centred about a negative value of $p_3$.  As reported in a recent work \cite{ChitraMomentum} this, indeed, is the case.  For the multi-sheeted pulse, as shown in Fig.~\ref{E_A_t_Sau_OmTau5_b0c0}(c), the electric field oscillates within the smooth envelope. In contrast to the single-sheeted field case, the vector potential is much suppressed  in the tail region of the field having oscillatory structure in the centre, see Fig.~\ref{E_A_t_Sau_OmTau5_b0c0}(b). Resulting $E(t)/\omega^2(\textbf{p},t)$ has a temporal profile close that of $E(t)$, except near the pulse centre. The QEPP region is consequently broader and the temporal profile of $|\Phi(\textbf{p}, t)|$  in this region is modified by the subcycle structures of the electric field. In this case too, the temporal profile of $|\Phi(\textbf{p},t)|$ in the QEPP stage is well explained  by that of $|E(t)|/\omega^2(\textbf{p},t)$. The rapid oscillation of $|\Phi(\textbf{p},t)|$ in the transient region in this case is governed by the oscillations in $E(t)$, $\cos{\psi(\textbf{p},t)}$ and $A(t)$, therefore the transient region is elongated and the modulation effect is seen in Fig.~\ref{F_t_Sau_OmTau_05}(b).
In the QEPP region, the counter term $v(\textbf{p},t)$ governing the depolarization/ pair annihilation is stronger than $u (\textbf{p},t)$  which is responsible for the polarization/ pair creation (Fig.~\ref{E_A_t_Sau_OmTau5_b0c0}(d)-(e)). Both $u(\textbf{p},t)$ and $v(\textbf{p},t)$ oscillate with varying amplitudes which is large in the centre of the pulse. The decrease in amplitude of $v(\textbf{p},t)$ in moving away from the centre is more than that of $u(\textbf{p},t)$. In the transient region both the amplitudes are nearly same, before becoming identical in the REPP stage.

\subsection{Effect of frequency chirping on field induced phase transition}
It is clear from the results discussed so far that the complexity in the evolution of the modulus and the phase of the order parameter, particularly in the transient stage, is because of the non linear coupling in the dynamical equations governing the evolution of the modulus and the phase. As demonstrated above, in the QEPP region, the evolution of modulus is mostly governed by both electric field $E(t)$ and vector potential $A(t)$. The evolution of the phase, on the other hand, is governed by two distinct terms which contain all the dynamic variables. In the QEPP stage, where the phase evolves slowly and smoothly, the two terms seem to balance each other. The transient stage arises when this dynamic balance is lost and hence there is steep increase in the phase over a very small duration. After the transient stage, the dynamics of the phase and modulus of the order parameter are decoupled, see Fig.~\ref{coslambdaOmT5}.
      
\begin{figure}[h]
\begin{center}
{\includegraphics[width = 2.5in]{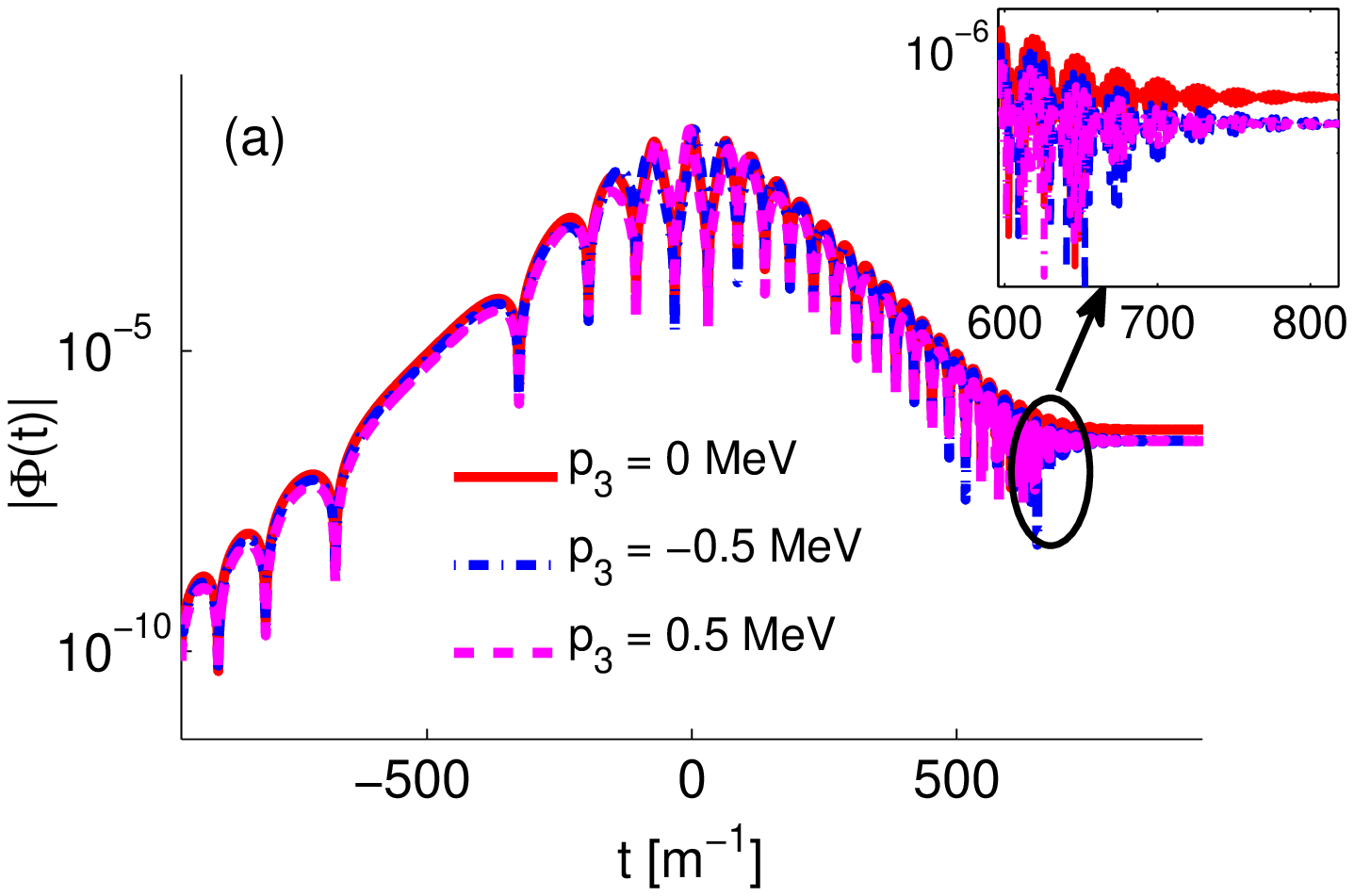} ~
\includegraphics[width = 2.5in]{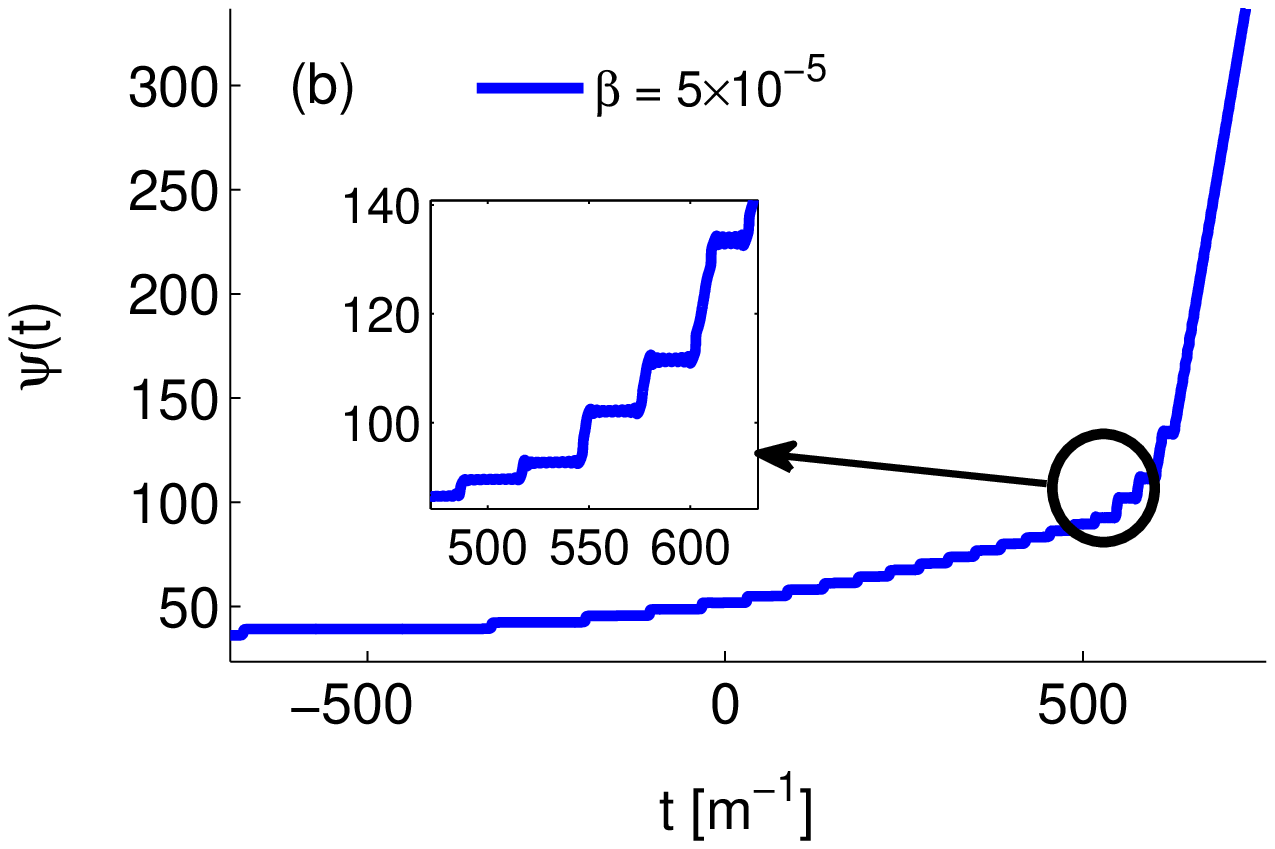} \\
\includegraphics[width = 2.5in]{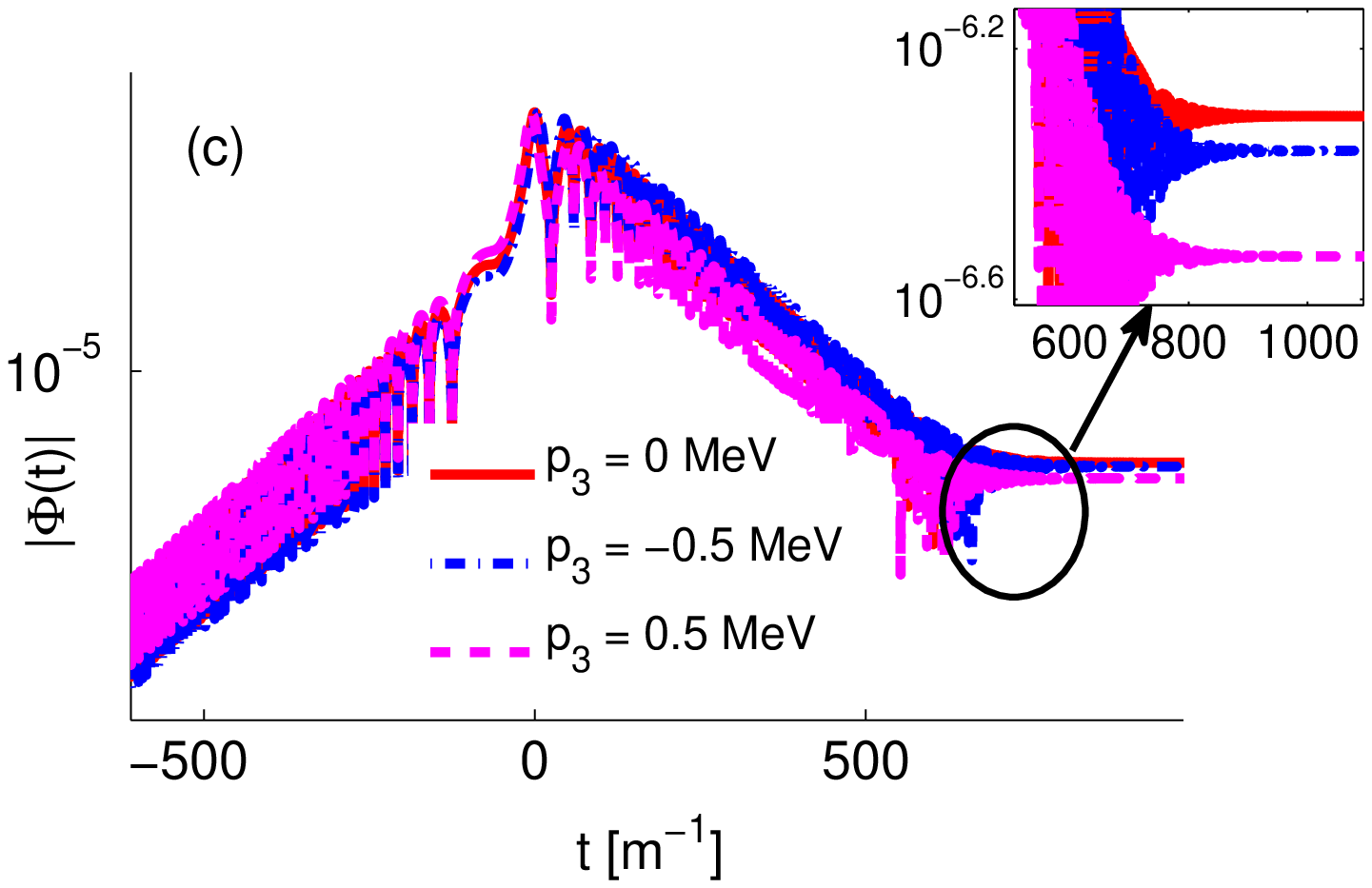} ~~
\includegraphics[width = 2.5in]{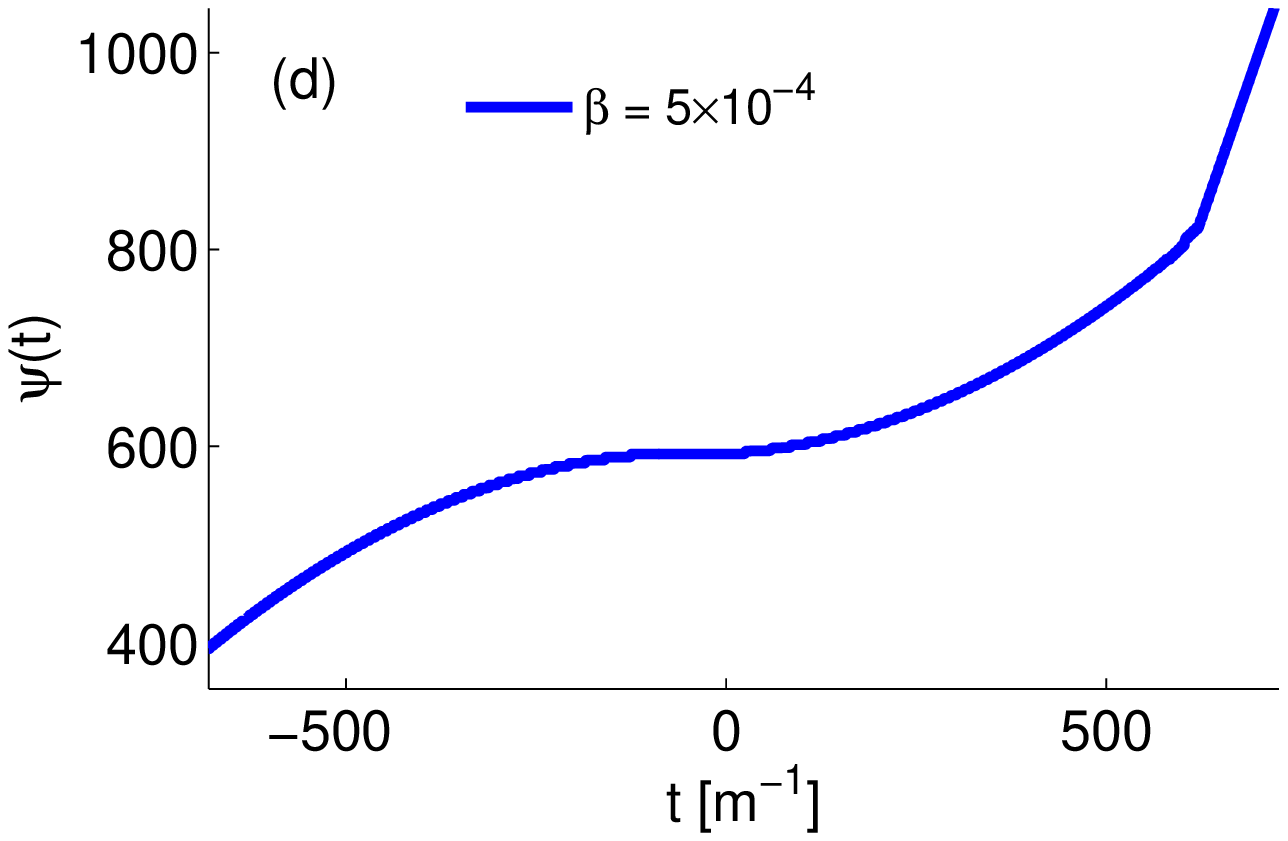}} 
\caption{Evolution of $|\Phi(\textbf{p}, t)|$ and $\psi(\textbf{p}, t)$ for time dependent multi-sheeted Sauter pulse  with linear frequency chirp. (a) and (b) $\beta = 5\times 10^{-5}$. (c) and (d) $\beta = 5\times 10^{-4}$. In (a) and (c) the longitudinal momentum $p_3 = 0, \pm 0.5$ MeV. In (b) and (d) $p_3 = 0$. The transverse momentum $p_{\perp}=0$ for all the cases. $\omega_0\tau = 5$. Other parameters are same as in Fig.~\ref{F_t_Sau_OmTau_05}.}
\label{F_t_Sau_cubic_b_c0}
\end{center}
\end{figure}

\begin{figure}[h]
\begin{center}
{\includegraphics[width = 2.0in]{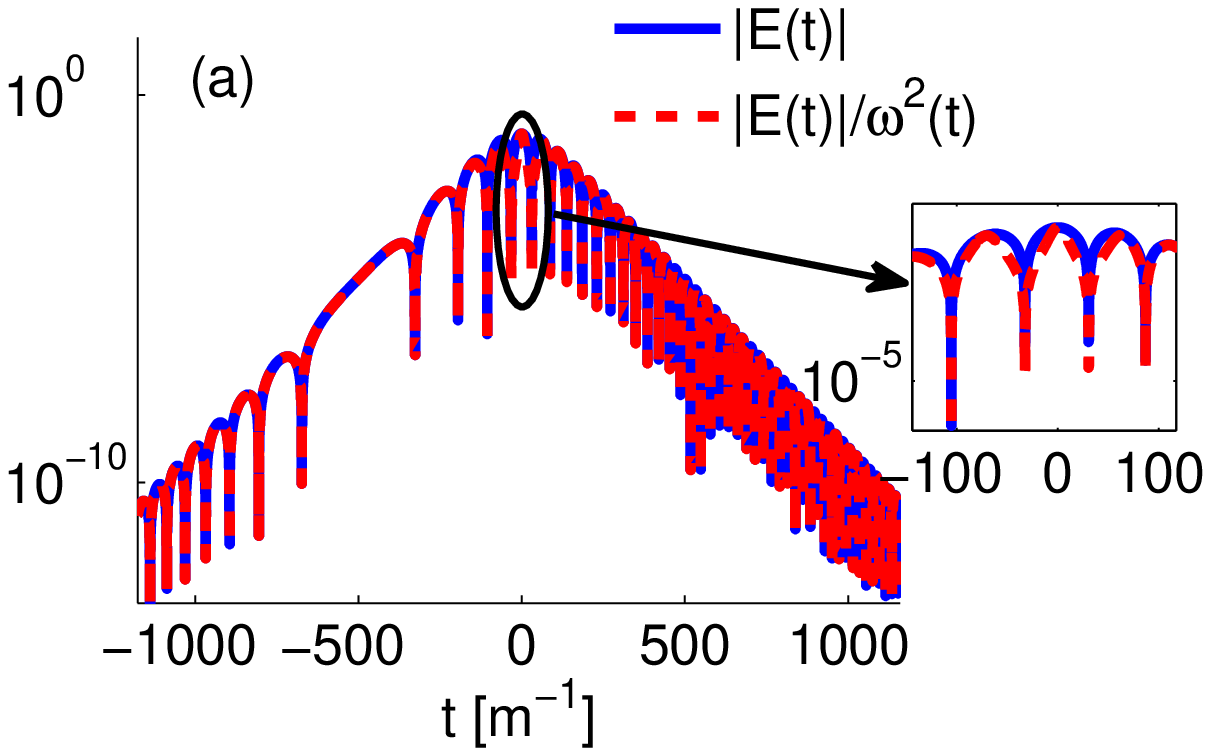}~
\includegraphics[width = 2.0in]{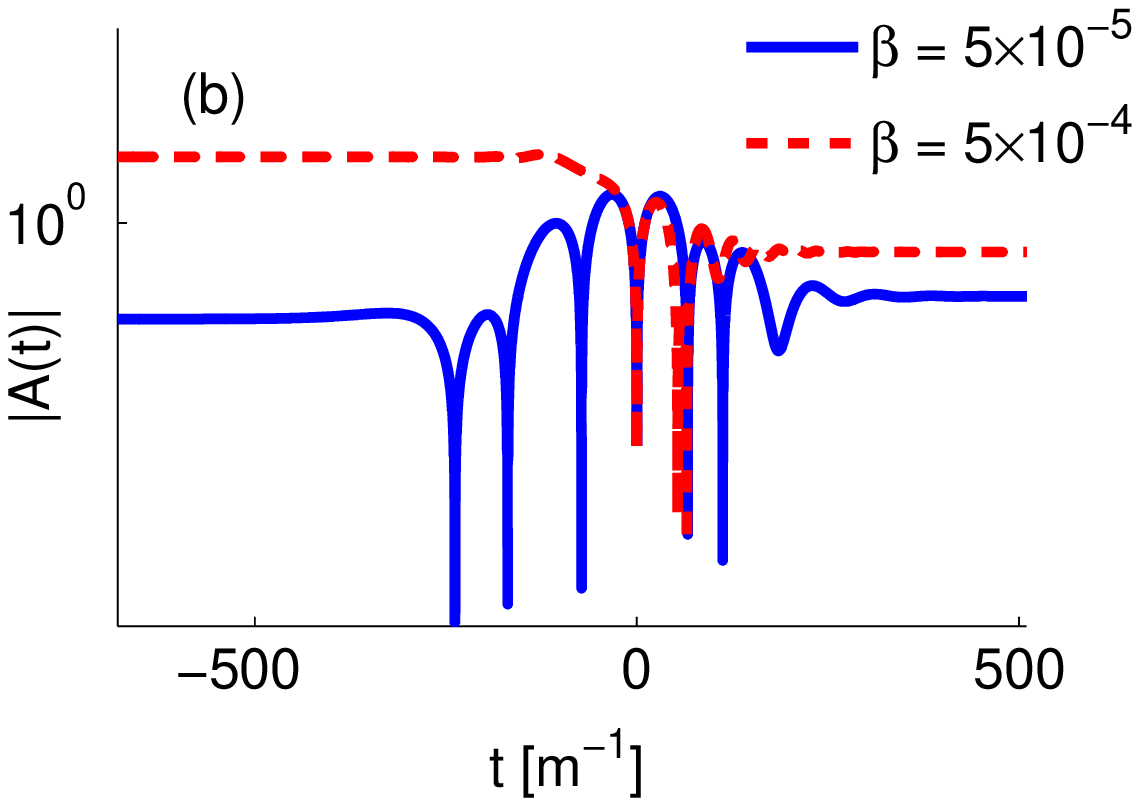}~
\includegraphics[width = 2.2in]{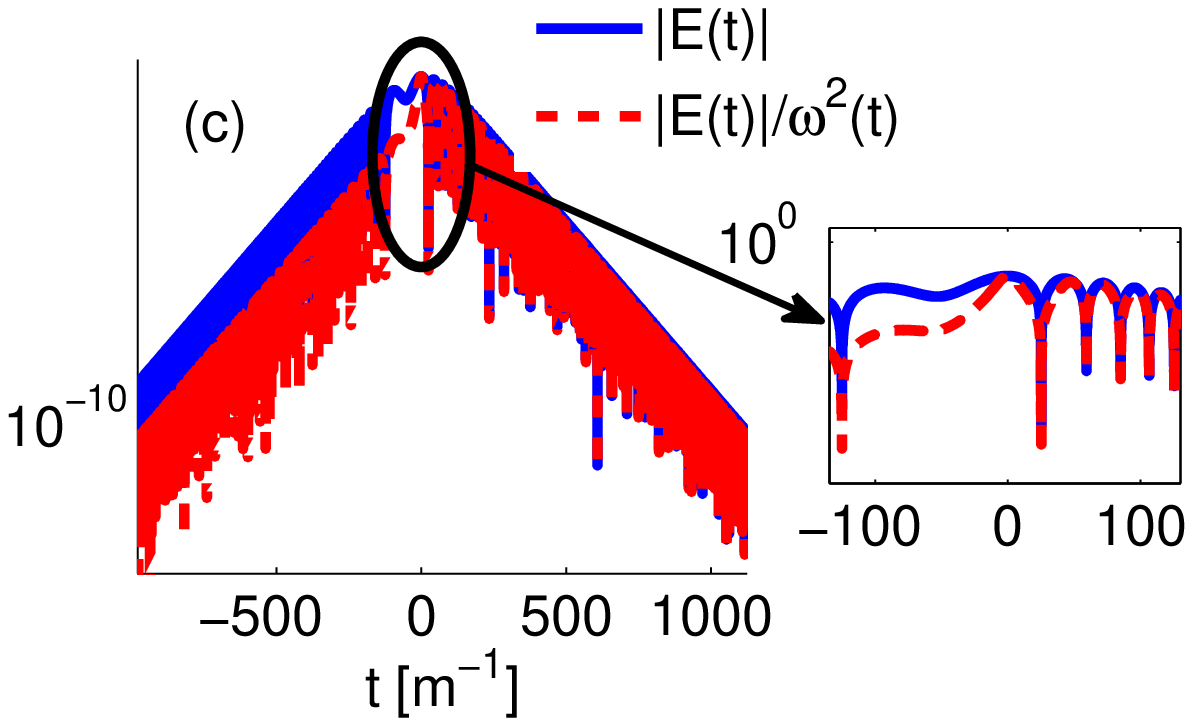}} 
\caption{ (a) and (c) $|E(t)|$ and $|E(t)|/\omega^2(\textbf{p}, t)$ for the two Sauter pulses used in Fig. \ref{F_t_Sau_cubic_b_c0} with $\beta = 5\times 10^{-5}$ and $\beta = 5\times 10^{-4}$, respectively. (b) $|A(t)|$  for both the Sauter pulses. Others parameters are same as in Fig.~\ref{E_A_t_Sau_OmTau5_b0c0}.}
\label{E_A_t_Sau_OmTb1_3c0}
\end{center}
\end{figure}

The presence of frequency chirping, in effect makes frequency time dependent. This, in turn  affects the number of subcycle oscillations within the envelope. In the presence of linear frequency chirp $\beta$, the subcycle oscillations are asymmetric about $t = 0$. As we have taken the positive value of $\beta$, the number of oscillations within the pulse envelope is lesser for $t < 0$ than that for $t > 0$. Hence the evolution of $|\Phi(\textbf{p},t)|$ shows irregular oscillations in the QEPP state (Fig.~\ref{F_t_Sau_cubic_b_c0}). As $\beta$ is increased further the oscillations become more rapid and are spread throughout the QEPP region as seen in Fig.~\ref{F_t_Sau_cubic_b_c0}(c). The evolution can qualitatively be understood by looking at the temporal  profiles of $E(t)$, $A(t)$  and also the ratio $|E(t)|/\omega^2(\textbf{p},t)$ as shown in Fig.~\ref{E_A_t_Sau_OmTb1_3c0}.
For $\beta = 5\times 10^{-5}$, the effect of frequency chirping is small. Hence the  evolution of $|\Phi(\textbf{p},t)|$, as in the case of multi sheeted Sauter pulse discussed above, follows the electric field profile in QEPP stage and the transient stage is marked by the sudden change in the evolution of the phase. For $\beta = 5\times 10^{-4}$, however, there is much  more asymmetry in the electric field profile and the vector potential is large and constant for $t<0$, before undergoing quick oscillations near $t=0$ and attaining a constant value thereafter. The evolution of $|\Phi(\textbf{p},t)|$ with enhanced oscillation frequencies in the QEPP stage follows the temporal profile of $|E(t)|/\omega^2(\textbf{p},t)$.  

\begin{figure}[h]
 \begin{center}
{ \includegraphics[width = 2.6in]{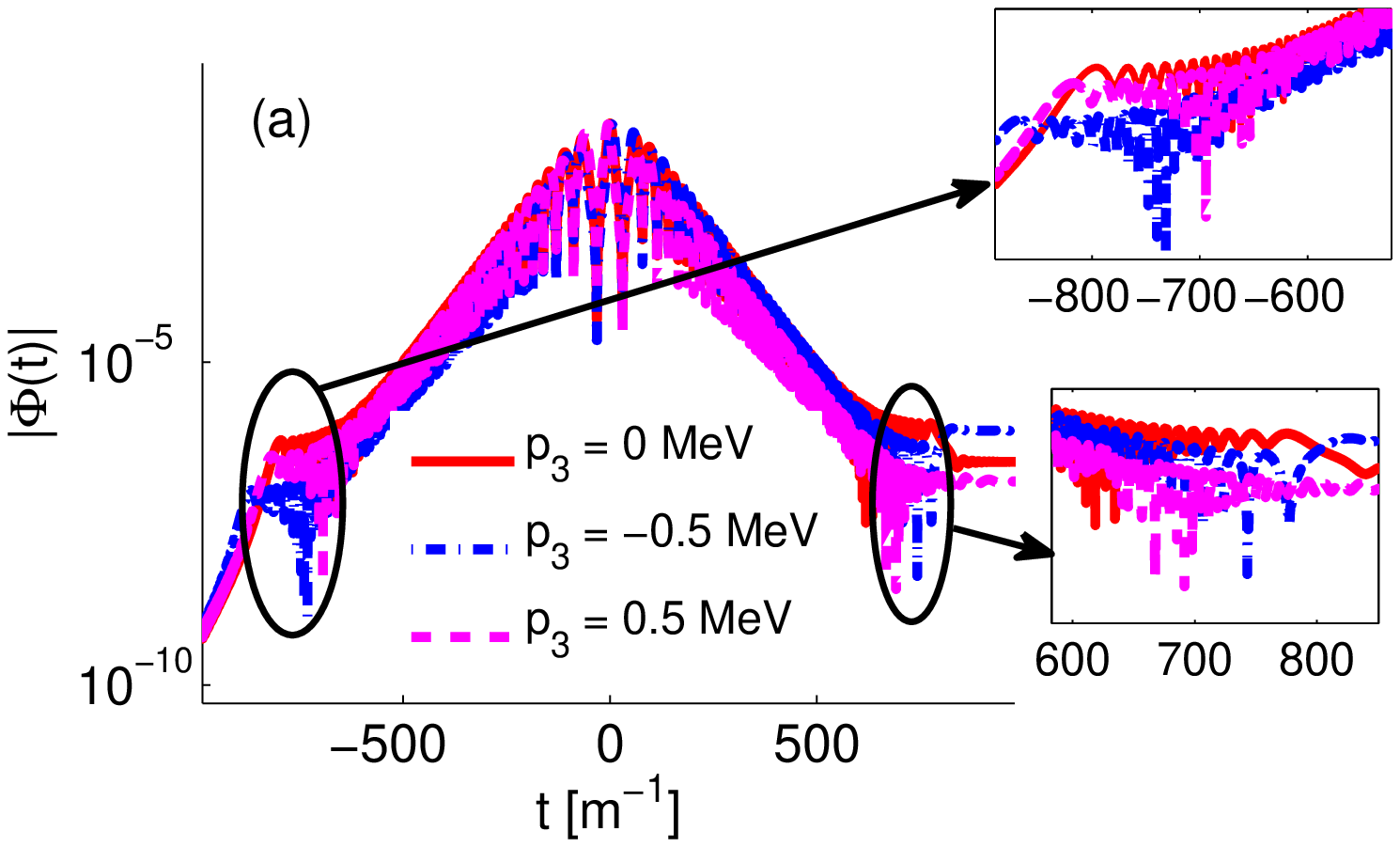}~
  \includegraphics[width = 2.6in]{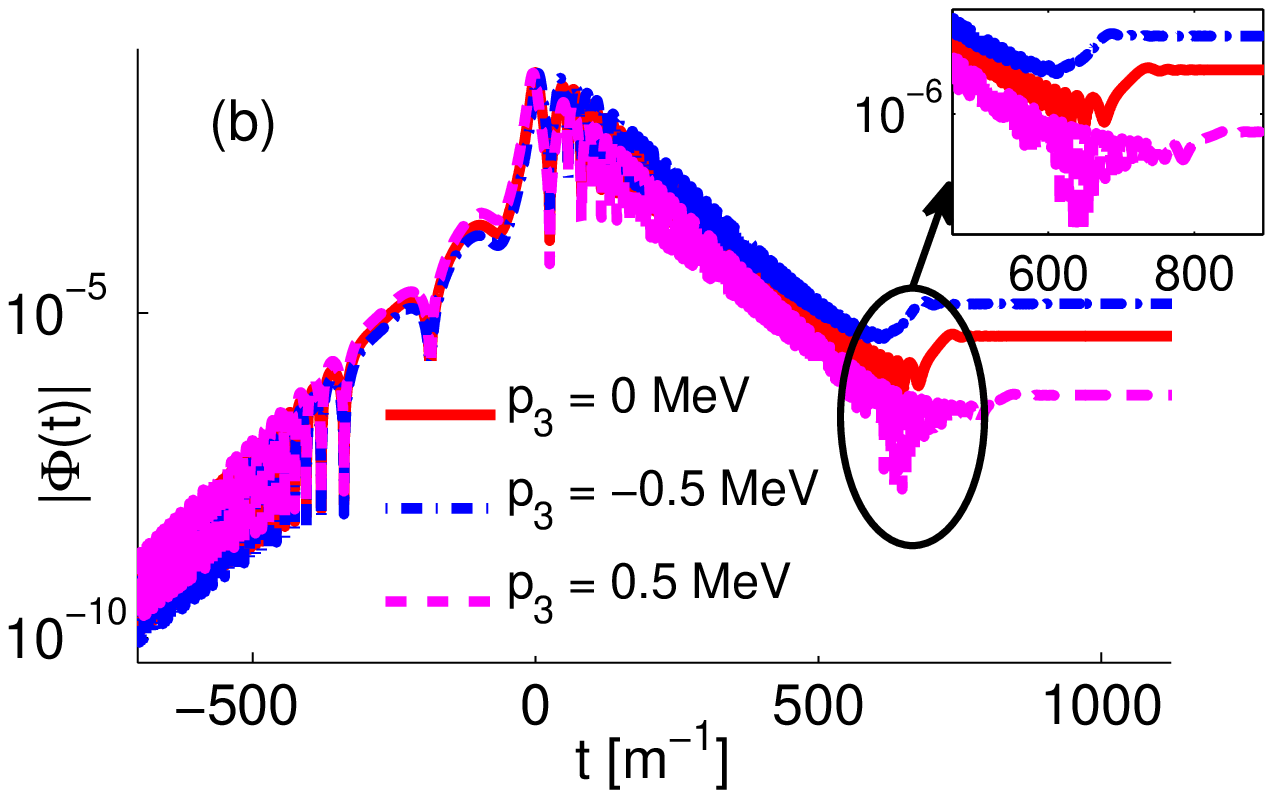} \\
   \includegraphics[width = 2.4in]{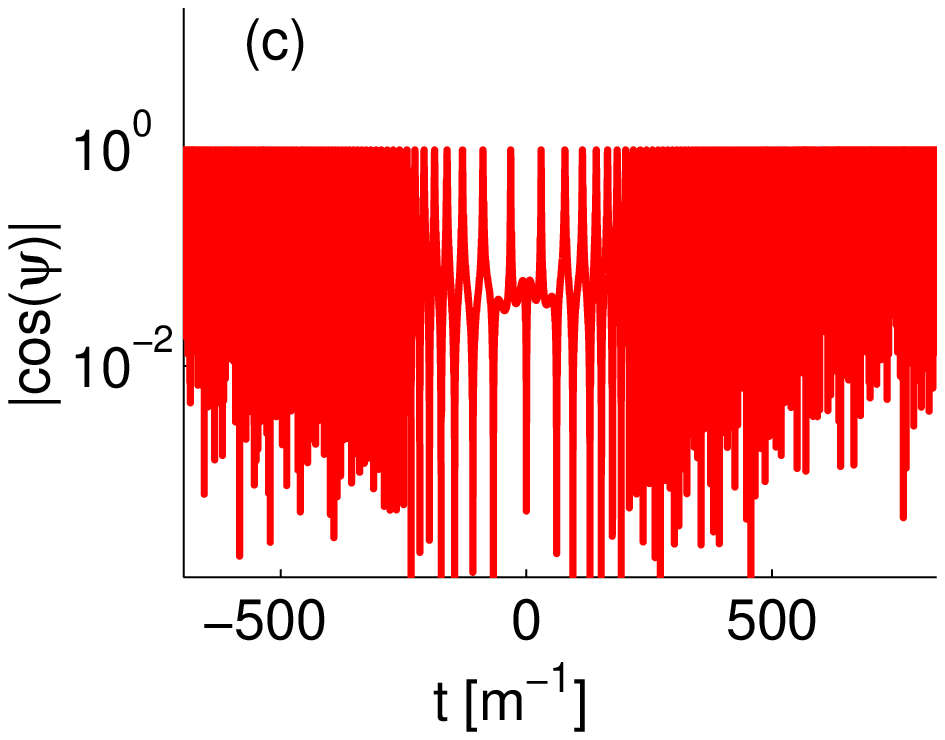} ~
    \includegraphics[width = 2.4in]{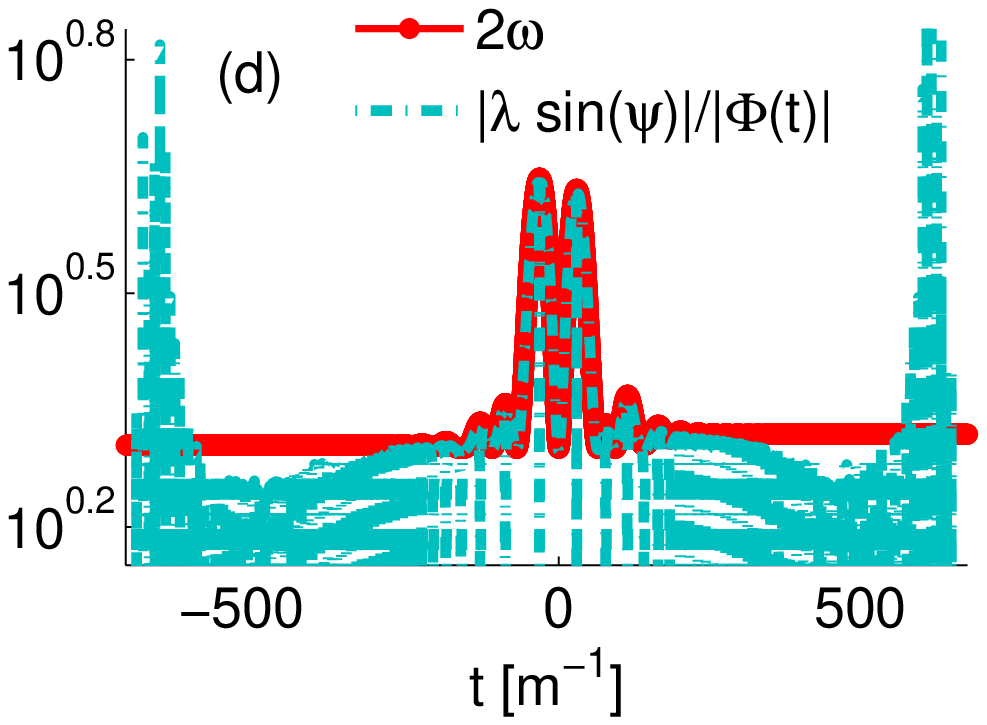}}  
 \caption{(a) Evolution of $|\Phi(\textbf{p}, t)|$ for a multi-sheeted Sauter pulse with linear chirp $\beta = 5\times 10^{-5}$, quadratic chirp $\alpha = 1\times 10^{-6}$ and the longitudinal momentum $p_3 = 0, \pm 0.5$ MeV. (b) Same as in (a) for a Sauter pulse with stronger linear frequency chirp  $\beta = 5\times 10^{-4}$. (c) $|\cos{\psi(\textbf{p}, t)}|$, (d) $2\omega(\textbf{p}, t)$ and $|\lambda\sin{\psi(\textbf{p}, t)}|/|\Phi(\textbf{p}, t)|$ for the Sauter pulse with $\beta = 5\times 10^{-4}$, $\alpha = 1\times 10^{-6}$ and for the longitudinal momentum $p_3 = 0$. $\omega_0\tau = 5$. The transverse momentum $p_{\perp}=0$ for all the cases. $\lambda = E(t)/\omega^2(\textbf{p}, t)$.  Other parameters are same as in Fig.~\ref{F_t_Sau_OmTau_05}.}
 \label{F_t_Sau_cubic_b_ci}
 \end{center}
 \end{figure}

We now study the effect of varying linear chirp parameter $\beta$ in the presence of a fixed quadratic chirp ($\alpha =1\times 10^{-6}$) on FIPT. For small values of $\beta$, the inequality $\omega_0 > \alpha\tau^2 >\beta\tau $ holds and the quadratic chirp dominates the evolution of $|\Phi(\textbf{p},t)|$. The evolution shows the formation of a pre-transient stage in the QEPP region for values of $t < 0$ (before the electric field reaches its maximum value $E_0$). This is shown in Fig.~\ref{F_t_Sau_cubic_b_ci}(a). The onset of the pre-transient region, like in the transient stage, is marked by the dominance of the $(E(t)/\omega^2(\textbf{p}, t))( \sin{\psi(\textbf{p}, t)}/|\Phi(\textbf{p}, t)|)$ term over the $2\omega(\textbf{p}, t)$  term in controlling the dynamics of $\psi(\textbf{p}, t)$ (Fig.~\ref{F_t_Sau_cubic_b_ci}(d)). This causes rapid oscillations in $\cos{\psi(\textbf{p}, t)}$ (Fig.~\ref{F_t_Sau_cubic_b_ci}(c))  and hence in $|\Phi(\textbf{p}, t)|$ (Fig.~\ref{F_t_Sau_cubic_b_ci}(a)) around the temporal region  near the pre-transient stage. 
 However the pre-transient stage gets suppressed for higher values of $\beta$ to give the uninterrupted QEPP region with irregular, fast and spread out oscillations as seen in Fig.~\ref{F_t_Sau_cubic_b_ci}(b). It is seen that for higher values of $\beta$, the formation of REPP stage takes place with larger magnitude of the order parameter for $p_3 = -0.5$ MeV mode than  $p_3 = 0.5$ MeV, suggesting thereby that the momentum spectrum pairs created by the pulse is centred at a negative value of longitudinal momentum. 
 
\begin{figure}[h]
 \begin{center}
{\includegraphics[width = 2.6in]{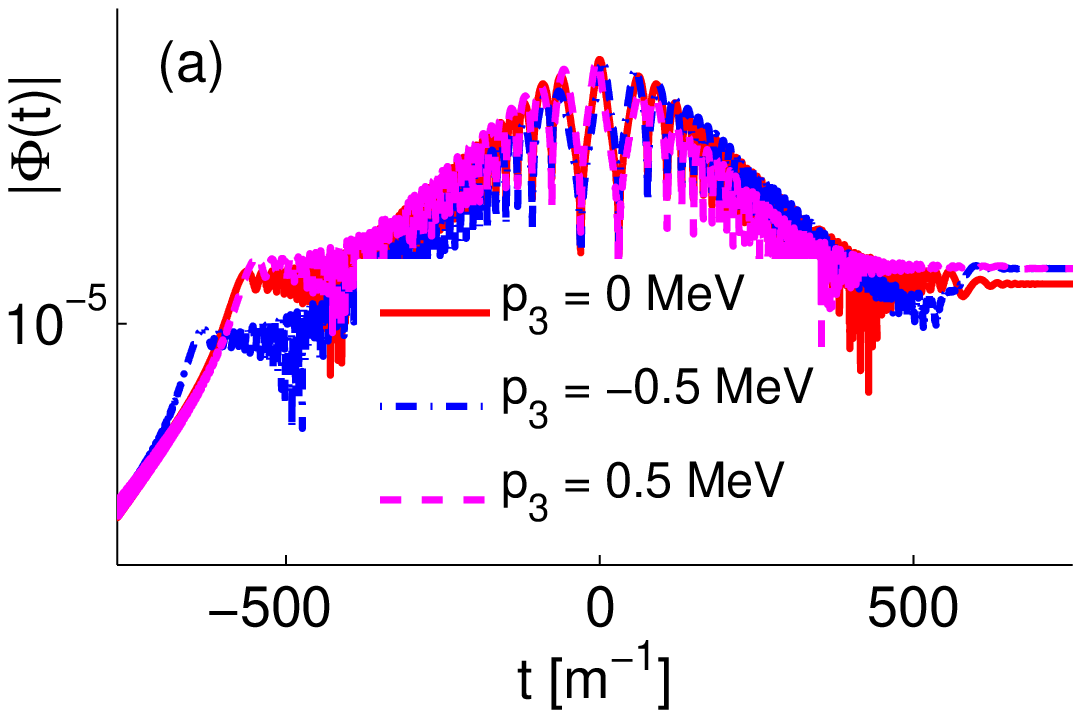}~~
\includegraphics[width = 2.6in]{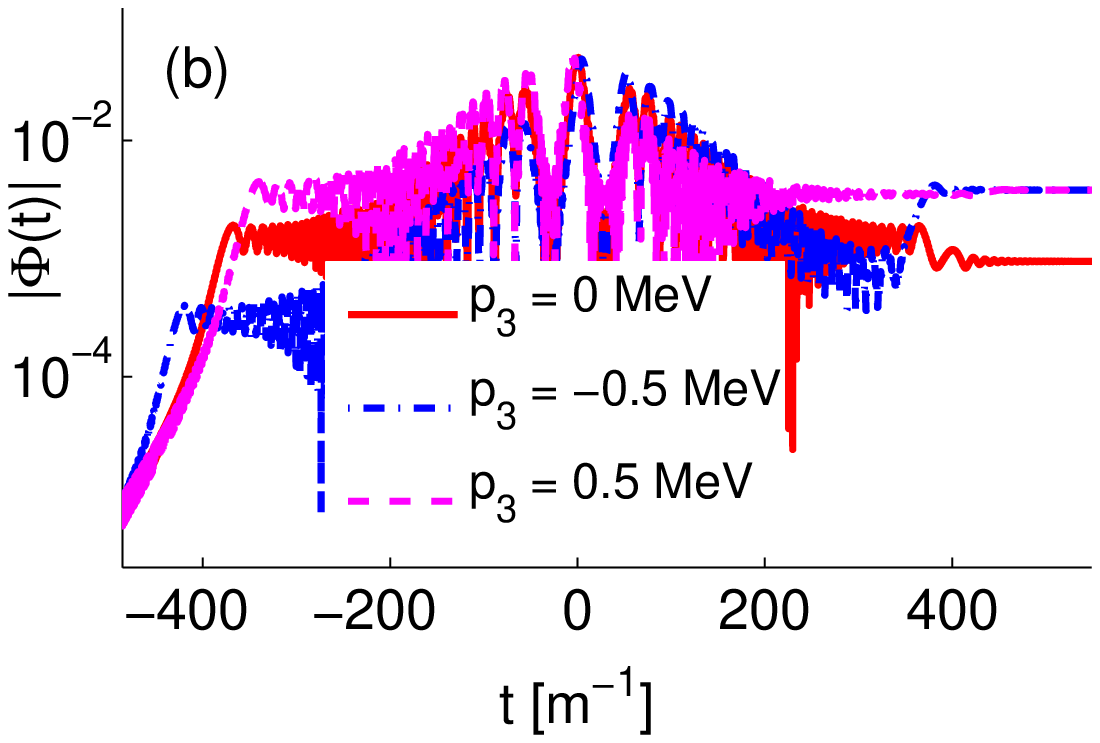}\\
\includegraphics[width = 2.3in]{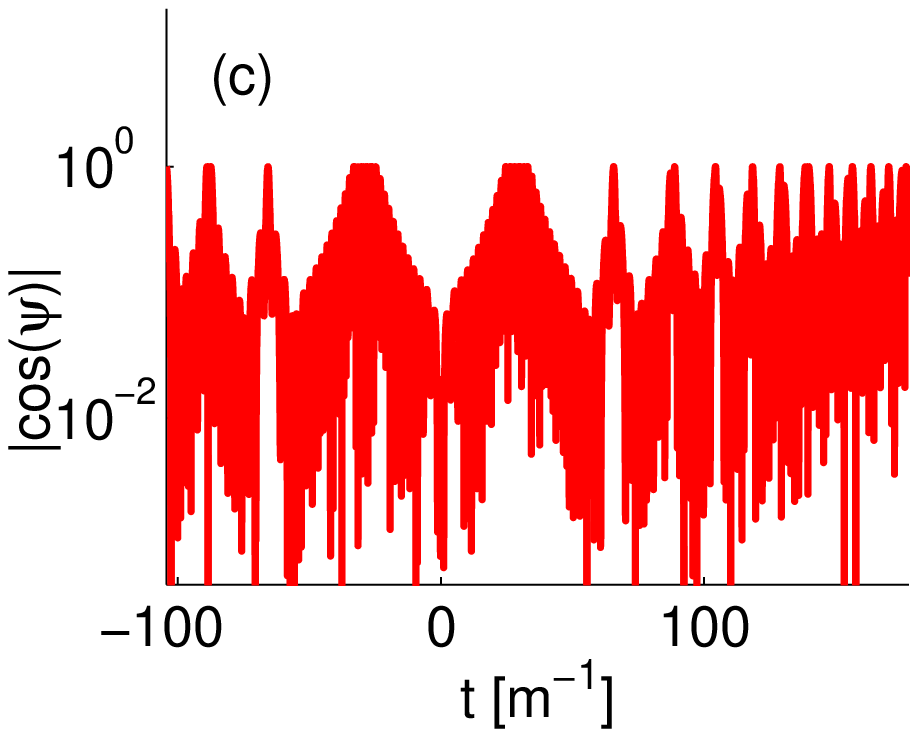}~~
\includegraphics[width = 2.3in]{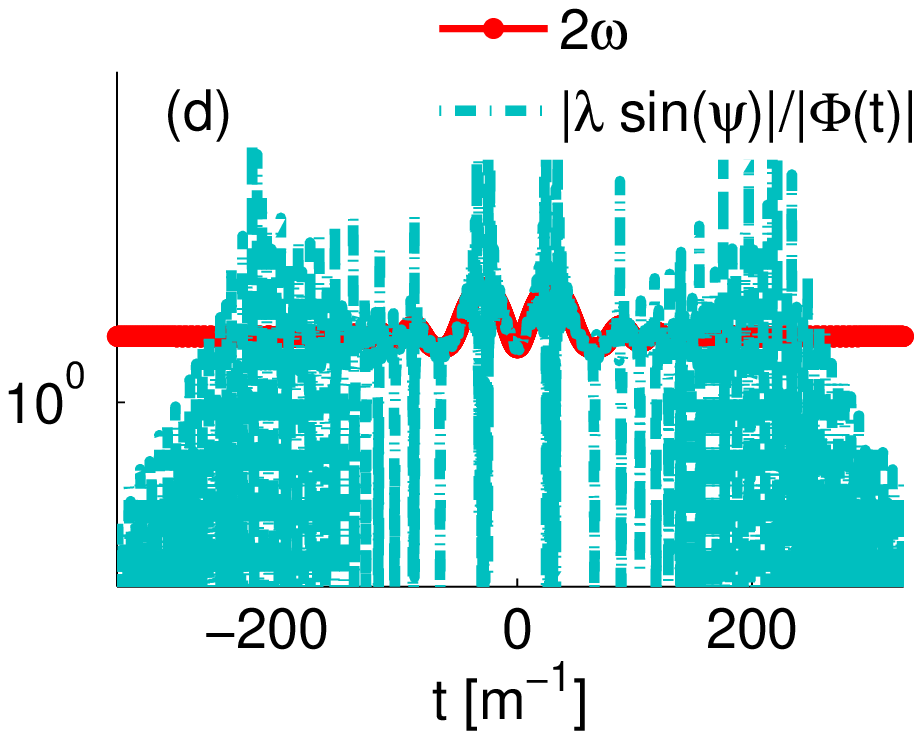}} 
 \caption{(a) $|\Phi(\textbf{p}, t)|$ for a  multi-sheeted Sauter pulse with only quadratic chirp ($\beta = 0$) with  $\alpha = 2\times 10^{-6}$. Longitudinal momentum $p_3 = 0, \pm 0.5$ MeV. (b) Same as in (a) for  $\alpha = 5\times 10^{-6}$. (c) $|\cos{\psi(\textbf{p}, t)}|$ (d) $2\omega(\textbf{p}, t)$ and $|\lambda\sin{\psi(\textbf{p}, t)}|/|\Phi(\textbf{p}, t)|$ for $\beta = 0$, $\alpha = 5\times 10^{-6}$ and $p_3 = 0$.  $\omega_0\tau = 5$. The transverse momentum $p_{\perp}=0$ for all the cases. $\lambda = E(t)/\omega^2(\textbf{p}, t)$. The other parameters are same as in Fig.~\ref{F_t_Sau_OmTau_05}.}
 \label{F_t_Sau_cubic_b0_c}
 \end{center}
 \end{figure}

 The effect of only quadratic chirping on FITP is shown in Fig.~\ref{F_t_Sau_cubic_b0_c} for the longitudinal momentum values $p_3 = 0$ MeV and $p_3 = \pm 0.5$ MeV. For $\alpha = 2\times 10^{-6}$ the evolution of $|\Phi(\textbf{p},t)|$ shows high frequency oscillations in  QEPP stage, as seen in Fig.~\ref{F_t_Sau_cubic_b0_c}(a). There is a formation of the pre-transient region in this case. When the value of $\alpha$ is increased to $\alpha = 5\times 10^{-6}$   the pre-transient and the transient  regions move closer to the maximum of the electric field and hence get closer to each other.  Higher the value of $\alpha$ larger is the shift. When the pre-transient and transient stages are close enough, as in Fig.~\ref{F_t_Sau_cubic_b0_c}(b) for example, the rapid oscillations in $\cos{\psi(\textbf{p}, t)}$ which set in at the pre-transient stage continue till the transient stage, see Fig.~\ref{F_t_Sau_cubic_b0_c}(c). This should be compared with the evolution of $\cos{\psi(\textbf{p}, t)}$ presented in Fig.~\ref{F_t_Sau_cubic_b_ci}(c) wherein the the rapid oscillations of a relatively far-off pre-transient stage do not extend all the way up to the transient stage. They are rather interrupted by slower oscillations in the central region of the pulse. An early occurrence of the transient stage leads to the formation of REPP stage at earlier times as the value of $\alpha$ is increased. Consequently, the modulus of the order parameter in the REPP region  increases with the increase in the value of $\alpha$. This results in the enhancement of pair production rate. Furthermore, a clear separation for different momentum modes is seen in the pre-transient and transient stages of evolution of order parameter.  
\section{conclusion}\label{concl}
To summarize, we have studied FIPT in presence of time dependent Sauter pulse by converting QKE for single particle distribution function into a set of two nonlinear coupled equation describing the evolution of modulus $|\Phi(\textbf{p}, t)|$ and phase $\psi(\textbf{p}, t)$ of the complex order parameter associated with the phase transition. Dynamics of $|\Phi(\textbf{p}, t)|$ is governed by the product $(E(t)/\omega^2(\textbf{p}, t))\cos{\psi(\textbf{p}, t)}$ and that of $\psi(\textbf{p}, t)$ is dictated by two competing terms in form of $2\omega(\textbf{p}, t)$, the dynamical energy gap,  and $(E(t)/\omega^2(\textbf{p}, t))(\sin{\psi(\textbf{p}, t)}/|\Phi(\textbf{p}, t)|)$, under the approximation $|\Phi(\textbf{p}, t)|\ll 1$ (which indeed is true for all the cases considered here). With the help of these coupled equations we have been able to shed light on the origin of different evolution stages reported in Refs.~\cite{Smolyansky2012TimeReversalSymmetry,Smolyansky2017FieldPhaseTrans,BlaschkeCPP} for the single-sheeted Sauter pulse and multi-sheeted Gaussian pulse. We find that in such cases $\psi(\textbf{p}, t)$ remains nearly constant or varies very slowly with time in the beginning as the competing terms governing the evolution nearly cancel out. Consequently, $\cos{\psi(\textbf{p}, t)}$ varies slowly with the time scale comparable to that of the external electric field, thereby resulting in the QEEP stage of FIPT wherein the evolution of $|\Phi(\textbf{p}, t)|$ follows the profile of $E(t)/\omega^2(\textbf{p}, t)$. The slow variation of $\psi(\textbf{p}, t)$ is followed by its abrupt steep increase. This makes $\cos{\psi(\textbf{p}, t)}$ oscillate much more rapidly than $E(t)/\omega^2(\textbf{p}, t)$. In this temporal regime, known as the transient stage evolution of $|\Phi(\textbf{p}, t)|$ is modulated by the rapid oscillation of $\cos{\psi(\textbf{p}, t)}$. With further increase in time, electric field strength becomes vanishingly small. As a result, the equations describing the dynamics of $|\Phi(\textbf{p}, t)|$ and $\psi(\textbf{p}, t)$ are decoupled. $|\Phi(\textbf{p}, t)|$ becomes a constant and the dynamics of $\psi(\textbf{p}, t)$ is governed by the dynamical energy gap $2\omega(\textbf{p}, t)$. 

$|\Phi(\textbf{p}, t)|$ is related to the single particle distribution function which gives number of created particles (antiparticles) for asymptotic times, which in this case is the REPP stage of evolution of the order parameter. In QEPP stage, the phase $\psi(\textbf{p}, t)$ gives a measure of coherence in the particle-antiparticle correlation function in the vacuum state. The oscillations of $\psi(\textbf{p}, t)$  in the transient stage of evolution signify ``dephasing'' which results in the loss of coherence of the correlation function which subsequently results in the creation of ``real'' pairs. The real part of the order parameter $u(\textbf{p}, t) = |\Phi(\textbf{p}, t)|\cos{\psi(\textbf{p}, t)}$ gives vacuum polarization, where as the imaginary part $v(\textbf{p}, t) = |\Phi(\textbf{p}, t)|\sin{\psi(\textbf{p}, t)}$ represent the corresponding counter term. It is found that in QEPP stage $u(\textbf{p}, t)\ll v(\textbf{p}, t)$. It is in the transient stage that $u(\textbf{p}, t)$ and $v(\textbf{p}, t)$ are nearly equal. Thus the process of pair production  is associated the sufficiently weakening of the ``counter term'' of the vacuum polarization.

In presence of frequency chirp, QEPP stage of evolution  of $|\Phi(\textbf{p}, t)|$ gets far more complex with faster and irregular oscillation which is attributed to the complexity in $E(t)/\omega^2(\textbf{p}, t)$. The most striking effect of quadratic frequency chirp is the appearance of a novel stage in the otherwise QEPP stage of the evolution of $|\Phi(\textbf{p}, t)|$, located symmetrically opposite to the transient stage from the pulse centre. The onset of the pre-transient region, as in the transient stage, is accompanied by the rapid oscillations in $\cos{\psi(\textbf{p}, t)}$, which in turn arises because of  dominance of the $(E(t)/\omega^2(\textbf{p}, t))( \sin{\psi(\textbf{p}, t)}/|\Phi(\textbf{p}, t)|)$ term over the $2\omega(\textbf{p}, t)$ term in the dynamics of $\psi(\textbf{p}, t)$.  With sufficient increase in the quadratic chirp, the pre-transient and transient stages move closer to the pulse centre. The rapid oscillations in $\cos{\psi(\textbf{p}, t)}$  continue uniterrupted from  the pre-transient to the transient stages. An early occurrence of the transient stage leads to the REPP stage with a higher value of $|\Phi(\textbf{p},\infty)|$ resulting thereby in the enhancement of the pair production rate. It may be possible to relate the rich dynamical features of the pre-transient and transient stages to the  vacuum polarization current which is in principle an experimentally measurable quantity.

Measuring pulse parameters of ultrashort and ultrintense laser pulses is a challenging task. Experiments based on the QED effects have been suggested to measure carrier envelope phase \cite{Dipiazza_CEP,ChitradipCEP}, relative content of e-and h-waves \cite{Banerjee2017}, relative phase \cite{ChitraPhase} and temporal envelope \cite{ChitraMomentum} of the counterpropagating pulses. Our studies suggest that enhancement of the pair production rate and polarization current can be used to measure the quadratic chirp parameter of ultrashort pulses.


\end{document}